\documentclass[10pt,emulateapj,apj,twocolumn]{emulateapj}
\usepackage{color}
\usepackage{graphics}     
\usepackage{import}
\usepackage{multirow}
\usepackage{booktabs}

\newcommand\fns{\footnotesize}
\shorttitle{Redshift survey of Abell 2199}
\shortauthors{Song et al.}
\begin{document}
\title{A REDSHIFT SURVEY OF THE NEARBY GALAXY CLUSTER ABELL 2199: 
  COMPARISON OF THE SPATIAL AND KINEMATIC DISTRIBUTIONS OF GALAXIES WITH THE INTRACLUSTER MEDIUM}
\author{Hyunmi Song\altaffilmark{1}, Ho Seong Hwang\altaffilmark{1}, Changbom Park\altaffilmark{2}, Takayuki Tamura\altaffilmark{3}}
\altaffiltext{1}{Quantum Universe Center, Korea Institute for Advanced Study, Heogiro 85, Dongdaemun-gu, Seoul 02455, Korea}
\altaffiltext{2}{School of Physics, Korea Institute for Advanced Study, Heogiro 85, Dongdaemun-gu, Seoul 02455, Korea}
\altaffiltext{3}{Institute of Space and Astronautical Science (ISAS), Japan Aerospace Exploration Agency (JAXA), Kanagawa 252-5210, Japan}
\begin{abstract}
We present the results from an extensive spectroscopic survey of the central region
  of the nearby galaxy cluster Abell 2199 at $z=0.03$.
By combining 775 new redshifts from the MMT/Hectospec observations
	with the data in the literature, we construct a large sample of 1624 galaxies 
	with measured redshifts at $R<30^\prime$,
	which results in high spectroscopic completeness at $r_\textrm{\fns petro,0}<20.5$ (77\%).
We use these data to study the kinematics and clustering of galaxies
	focusing on the comparison with those of the intracluster medium (ICM) 
	from Suzaku X-ray observations.
We identify 406 member galaxies of A2199 at $R<30^\prime$ using the caustic technique.
The velocity dispersion profile of cluster members appears 
	smoothly connected to the stellar velocity dispersion profile 
	of the cD galaxy.
The luminosity function is well fitted with a Schechter function at $M_r<-15$.
The radial velocities of cluster galaxies generally 
	agree well with those of the ICM,
but there are some regions where the velocity difference between the two
	is about a few hundred kilometer per second.
The cluster galaxies show a hint of global rotation
  at $R<5^\prime$ with $v_\textrm{\fns rot}=300{-}600\,\textrm{km s}^{-1}$,
  but the ICM in the same region do not show such rotation.
We apply a friends-of-friends algorithm to the cluster galaxy sample at $R<60^\prime$
	and identify 32 group candidates,
	and examine the spatial correlation between the galaxy groups and X-ray emission.
This extensive survey in the central region of A2199
	provides an important basis for future studies of interplay 
	among the galaxies, the ICM and the dark matter in the cluster.
\end{abstract}

\section{INTRODUCTION}
Galaxy clusters are the largest gravitationally bound objects in the universe.
They generally form at the nodes of filaments in the large-scale structure of the universe,
	and have grown through continuous accretion of material from the surroundings \citep{Kravtsov_Borgani2012}.
They consist of three components that include dark matter,
	the intracluster medium (ICM), and galaxies.
Each component has a different mass fraction in a cluster
	(dark matter: 80--95\% , ICM: 5--20\%, galaxies: 0.5--3\%; \citealt{Lin_etal2003,Ettori_etal2009}),
	making up the total mass of $10^{14}{-}10^{15} M_\odot$ \citep[e.g.][]{Rines_etal2013}.

The three components interact with each other as clusters evolve through cosmic time
	\citep{PH2009,Gu_etal2013}. 
Because of different physical properties of the three components,
	the evolution of each component within a cluster is diverse.
For example, when a galaxy cluster interacts or merges with another cluster,
	the three components can behave differently.
The measurement of spatial offsets among the three components
	can provide strong constraints on models regarding nature of dark matter
	\citep[e.g.][]{Markevitch_etal2004,Harvey_etal2015}.

Different kinematic properties of the collisional ICM and collisionless cluster galaxies also
	provide an important hint of cluster merging history.
Some numerical studies have suggested that
	off-axis merging between two clusters provides angular momentum to clusters
	\citep{Ricker1998,Takizawa2000,Ricker_Sarazin2001}
	and the resulting bulk motion of the ICM survives longer than that of galaxies \citep{Roettiger_Flores2000}.
Thus, a comparison of bulk motions between the ICM and galaxies allows us
	to infer when the clusters have experienced the major mergers.
Moreover, the relevant bulk motion in clusters can contribute to the kinetic Sunyaev-Zeldovich
	signals that can affect the cosmic microwave background analysis \citep{Dupke_Bregman2002}.
It is therefore necessary to examine the spatial distributions and the kinematics 
	of the ICM and galaxies in clusters to better understand the formation and evolution 
	of galaxy clusters and their constituents.

A2199 is one of nearby, rich, and X-ray bright clusters at $z=0.03$.
It hosts a massive cD galaxy (NGC 6166) at the center, 
	which shows a radio jet (3C 338) 
	interacting with surrounding material \citep{Nulsen_etal2013}.
A2199 forms a supercluster with neighboring clusters and groups 
	that are probably infalling into A2199 \citep{Rines_etal2001}.
Because of proximity and of a wealth of structures with different scales,
	A2199 has been an ideal laboratory to test structure formation models, 
	in particular the evolution of galaxies in connection to the ICM and dark matter.

There are also many multiwavelength surveys 
	that uniformly covered this supercluster region
	including the optical photometric and spectroscopic data from 
	Sloan Digital Sky Survey \citep[SDSS,][]{York_etal2000},
	Wide-field Infrared Survey Explorer \citep[WISE,][]{Wright_etal2010},
	mid-infrared photometric data, 
	and ROSAT and Suzaku X-ray data (\citealt{Voges_etal1999}, T. Tamura et al. 2017, in prep.).
The combination of these data provides interesting insights
	on galaxy properties and their environmental dependence in the supercluster region
	\citep[e.g.][]{Rines_etal2001,Rines_etal2002,Hwang_etal2012,LeeGH_etal2015}.

Here, we report the results from our extensive redshift survey of the central region of A2199
	at $R<30^\prime$ ($\sim$half virial radius of A2199).
We increase the spectroscopic completeness in the survey region 
	significantly from 38\% to 77\% at $r_\textrm{\fns petro,0}<20.5$,
	and construct a large sample of galaxies by combining with the data in the literature.
We use this extensive data set to compare the kinematics and the spatial distribution 
	of galaxies with those of the ICM.

This paper is constructed as follow. 
Section 2 describes the redshift data obtained with our MMT/Hectospec observations.
We determine cluster membership,
	and derive the velocity dispersion profile and the luminosity function in Section 3.
In Section 4, we compare the kinematics and the spatial distributions between galaxies and the ICM.
Unless explicitly noted otherwise, we adopt flat $\Lambda$-CDM cosmological parameters of
	$H_0=100\,h \textrm{km s}^{-1} \textrm{Mpc}$, $\Omega_\Lambda=0.7$, and $\Omega_M=0.3$.

\section{DATA}
\subsection{Photometric Data}
We used the photometric data of the SDSS data release 12 \citep[DR12;][]{SDSSDR12}
	to select targets for spectroscopic observations.
Because our goal is to conduct a complete, uniform redshift survey 
	of the central region of A2199 at $R<30^\prime$, we chose the galaxies at
	$r_\textrm{\fns petro,0}<21$ without using any color selection criteria
	\footnote{The subscript 0 of magnitudes represents magnitudes after the galactic extinction correction.}
	(i.e. extended sources in SDSS; see Section 4.2 of \citealt{Strauss_etal2002} for the star-galaxy separation in SDSS).
We weighted the spectroscopic targets with their apparent magnitudes.

\subsection{Spectroscopy}
We used Hectospec on the MMT 6.5m telescope for spectroscopic observations
	\citep{Fabricant_etal1998,Fabricant_etal2005}.
Hectospec is a 300-fiber multi-object spectrograph with a circular field of view (FOV) of a $1^\circ$ diameter.
We used the 270 line mm$^{-1}$ grating of Hectospec
	that provides a dispersion of 1.2$\AA$ pixel$^{-1}$ and a resolution of $\sim$6$\AA$.
We observed 4 fields with 3$\times$20 minutes exposure each,
	and obtained spectra covering the wavelength range 3500--9150$\AA$.
All the fields are centered on the A2199 X-ray center
	(R.A.$=247^\circ.1582$, decl.$=39^\circ.5487$; \citealt{Bohringer_etal2000}).

We reduced the Hectospec spectra with HSRED v2.0 
	that is an updated reduction pipeline originally developed by Richard Cool.
We then used RVSAO \citep{Kurtz_Mink1998} to determine the redshifts
	by cross-correlating the spectra with templates.
RVSAO gives the \citet{Tonry_Davis1979}'s $r$-value for each spectrum
	that is an indicator of cross-correlation reliability;
	we select only those galaxies with $r>4$,
	consistent with the limit confirmed by visual inspection
	\citep{SHELS2014,SHELS2016}.
In the end, we obtain 775 reliable redshifts from the observations of 1024 targets. 
We combine these data with those from the SDSS DR12 \citep{SDSSDR12}
	and from the literature \citep{RG2008,Rines_etal2002,Smith_etal2004}.
In total, we have 1625 redshifts at $R<30^\prime$;
	775 from this study, 477 from \citet{RG2008}, 363 from the SDSS DR12,
	and 10 from other studies.

\begin{figure*}
\centering
\includegraphics[width=0.7\textwidth]{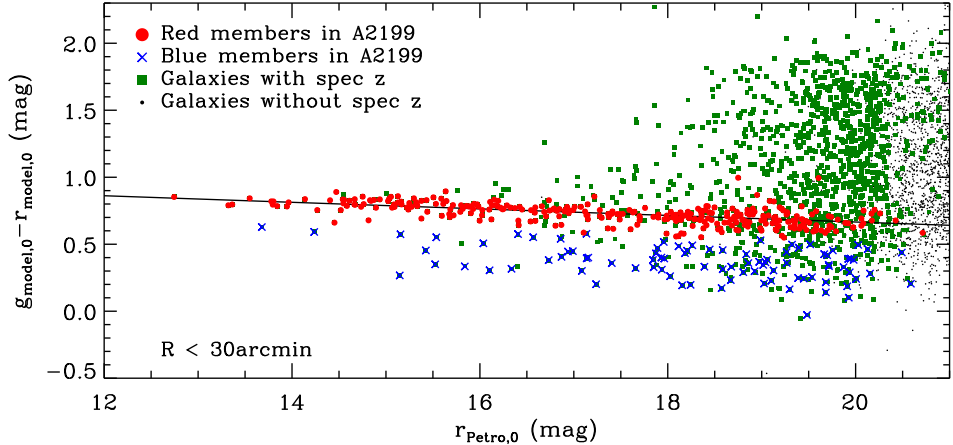}
\caption{
$(g_\textrm{\fns model,0}-r_\textrm{\fns model,0}){-}r_\textrm{\fns petro,0}$ 
color-magnitude diagram of A2199 galaxies at $R<30^\prime$.
Green dots represent galaxies with measured redshifts and black dots represent galaxies without redshifts.
Red dots denote cluster members which are in the red sequence, and blue crosses are blue cluster members.
}
\label{fig_ccmr}
\end{figure*}

Figure \ref{fig_ccmr} shows the ($g-r$)-$r$ color-magnitude diagram 
	for the objects in the central region of A2199 at $R<30^\prime$.
Galaxies with measured redshifts are represented with symbols in colors, 
	while those without spectra are indicated with black dots.
The red sequence of A2199 is clearly visible, 
	and the blue population is sparsely distributed below the red sequence.
We use the cluster member galaxies at $R<30^\prime$ (see Section 3.1 for the member selection) 
	to determine the best-fit relation of the red sequence, which is
\begin{equation}
   g_\textrm{\fns model,0}-r_\textrm{\fns model,0} 
  = 1.156 - 0.024 \, r_\textrm{\fns petro,0}
\label{eqn_redseq}
\end{equation}
	(solid line in Figure \ref{fig_ccmr}).
The rms scatter around this relation is 0.05 mag.
We divide the cluster member galaxies into two subsamples (red and blue)
	using the line 3$\sigma$ blueward of the best-fit red-sequence relation
	\citep{SanchezBlazquez_etal2009,Rines_etal2013,Hwang_etal2014};
	they are indicated separately with red dots and blue crosses in Figure \ref{fig_ccmr}.
Green squares are background plus foreground galaxies.

\begin{figure*}
\centering
\includegraphics[width=0.7\textwidth]{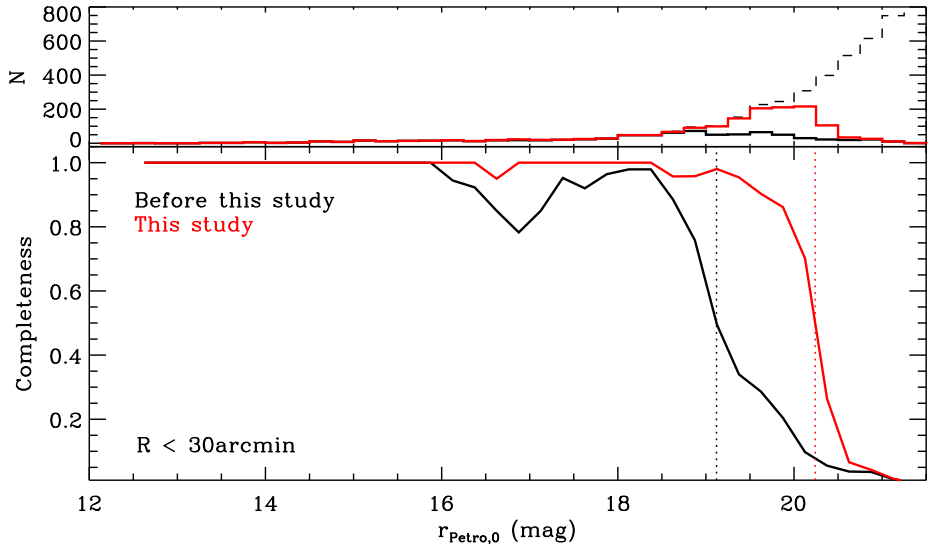}
\caption{(Top) Numbers of 
galaxies at $R<30^\prime$ as a function of $r_\textrm{\fns petro,0}$:
dashed line -- all the galaxies regardless of redshift measurement,
black solid line -- galaxies with measured redshifts before our MMT/Hectospec observations,
red solid line -- galaxies with measured redshifts after our MMT/Hectospec observations.
(Bottom) Differential spectroscopic completeness at $R<30^\prime$
as a function of $r_\textrm{\fns petro,0}$.
Black and red solid lines represent the completeness curves 
before and after our MMT/Hectospec observations, respectively.
Two vertical dotted lines denote apparent magnitudes
when the completeness drops below 50\%:
black dotted line -- $r_\textrm{\fns petro,0}=19.1$,
red dotted line -- $r_\textrm{\fns petro,0}=20.2$.
}
\label{fig_scompl_r}
\end{figure*}

\begin{figure*}
\centering
\includegraphics[width=0.68\textwidth]{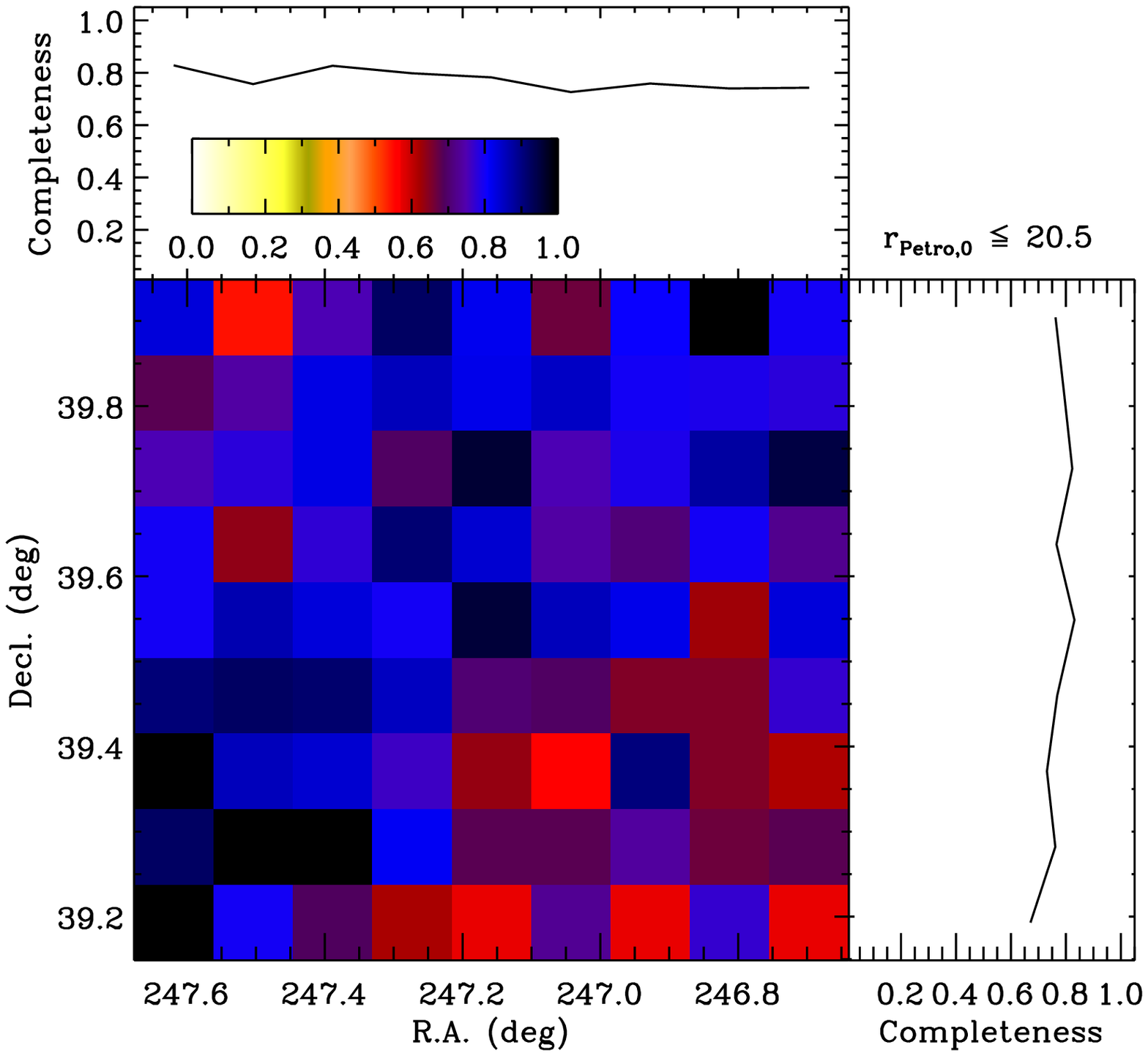}
\caption{Two-dimensional, cumulative spectroscopic completeness in the field of A2199
as a function of right ascension and of declination (bottom left).
Integrated spectroscopic completeness as a function of right ascension (top)
and declination (bottom right).
}
\label{fig_scompl_radec}
\end{figure*}

The bottom panel of Figure \ref{fig_scompl_r} shows 
	the spectroscopic completeness at $R<30^\prime$
	(i.e. the number of galaxies with measured redshifts divided 
	by the number of galaxies in the photometric sample)
	as a function of $r$-band apparent magnitude;
	black and red lines indicate the completeness before and after
	our Hectospec observations, respectively.
The vertical lines indicate the magnitudes 
	when the differential completeness drops below 50\%.
The effective magnitude limit including our new redshift data
	is fainter than the one based on previous redshift data by about one magnitude (from 19.1 to 20.2).
Similarly, the cumulative completeness at $r_\textrm{\fns petro,0}<20.5$ has
	significantly increased from 38\% to 77\% with our new data.
The bottom left panel of Figure \ref{fig_scompl_radec} 
	shows the two-dimensional map of the spectroscopic completeness
	at $r_\textrm{\fns petro,0}<20.5$ as a function of R.A. and declination.
The top and right panels show the integrated completeness as a function of R.A.
	and declination, respectively, showing very little variation.
The high completeness with little spatial variation
	suggests that the cluster is very successfully and uniformly covered by our redshift survey.

\begin{figure}
\centering
\includegraphics[width=0.5\textwidth]{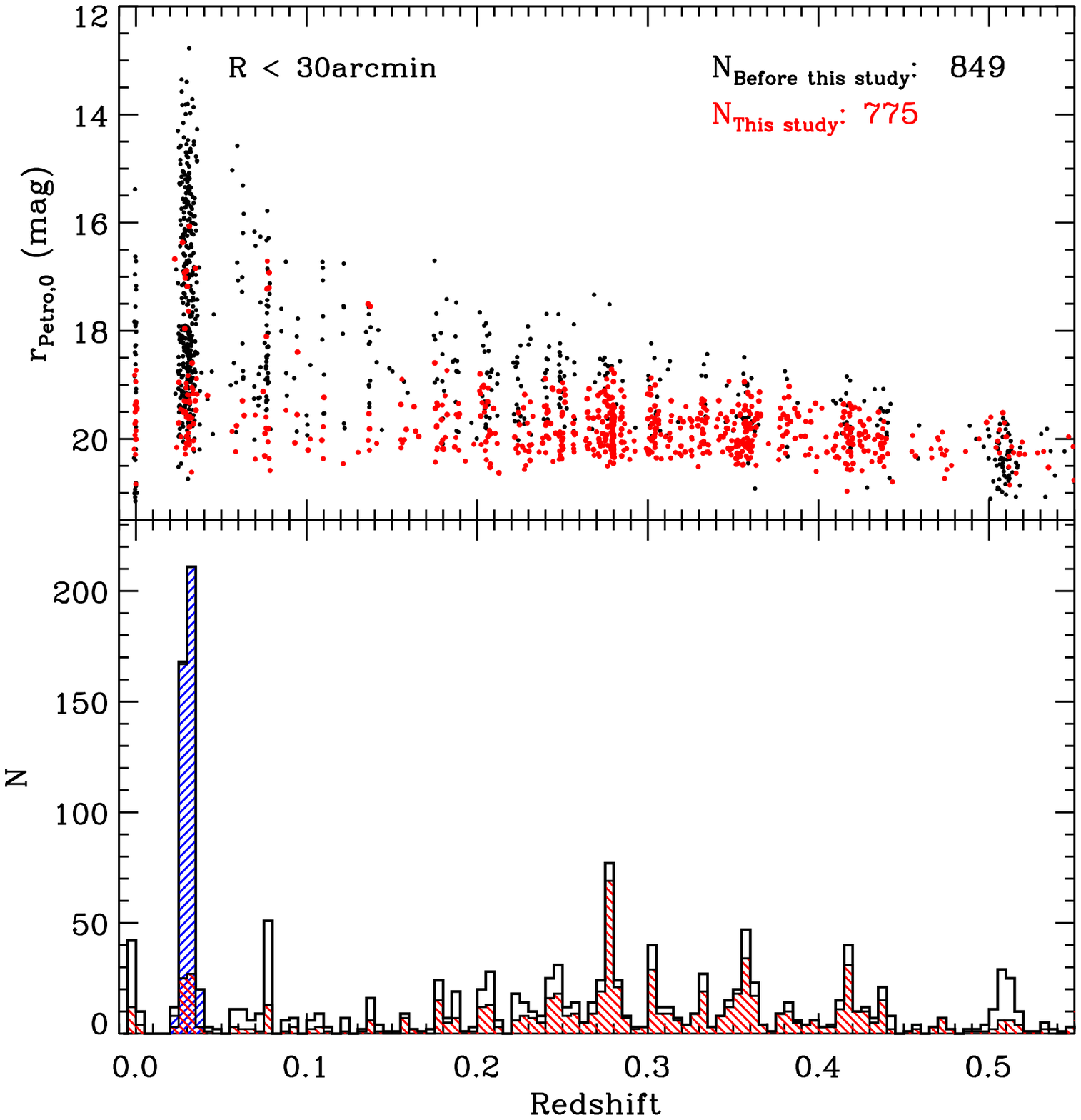}
\caption{(Top) Distribution of $r_\textrm{\fns petro,0}$ 
of galaxies at $R<30^\prime$ as a function of redshift.
Black dots represent galaxies with measured redshifts in the literature.
Red dots represent galaxies with new redshifts in this study.
(Bottom) Redshift distribution of galaxies.
Black and red colors indicate all the galaxies with measured redshifts at $R<30^\prime$
and the galaxies with new redshifts in this study, respectively.
Blue histogram represents the redshift distribution of the member galaxies of A2199.
}
\label{fig_zhist}
\end{figure}

The top panel of Figure \ref{fig_zhist} shows $r$-band apparent magnitudes of galaxies
	as a function of redshift. Red and black dots indicate the galaxies with measured redshifts
	from this study and from the literature, respectively.
As expected, we targeted mainly faint objects at $r_\textrm{\fns petro,0}=19.0{-}20.5$.
The bottom panel is for redshift histograms;
	red is for the new redshifts and black is for the total.
The blue histogram is for the A2199 member galaxies, which is peaked at $z=0.03$

In Table \ref{tab_samp},
	we list all the redshifts in the central region of A2199 at $R<1^\circ$.
In total, there are 2400 redshifts.
The table includes SDSS DR12 ObjID, R.A., declination, $r$-band apparent magnitude ($r_\textrm{\fns Petro,0}$),
	flag whether the source is an extended source based on the probPSF parameter in the SDSS database,
	redshift ($z$) and its error, redshift source, cluster membership (see Section 3.1),
	and group ID (see Section 5.1).

\begin{deluxetable*}{rccccccccc}
\tabletypesize{\footnotesize}
\tablewidth{0pc}
\tablecaption{Redshifts in the field of A2199 within $1^\circ$ from the cluster center}
\tablehead{
\multirow{2}{*}{ID} & \multirow{2}{*}{SDSS DR12 ObjID} & R.A. & Decl. & $r_{\rm Petro,0}$ & \multirow{2}{*}{Flag\tablenotemark{a}} & \multirow{2}{*}{$z$} & \multirow{2}{*}{$z$ Source\tablenotemark{b}} & \multirow{2}{*}{Member\tablenotemark{c}} & \multirow{2}{*}{Group ID\tablenotemark{d}} \\ & & ($^{\circ}$) & ($^{\circ}$) & (mag) & & & & & 
}
\startdata
   1 &    1237665355085054298 & 245.871175 &  39.650101 &  18.219 &  1 & $0.27797\pm0.00004$ &  2 &  0 &  0 \\
   2 &    1237659324955362134 & 245.881278 &  39.458171 &  19.350 &  0 & $0.38281\pm0.00010$ &  2 &  0 &  0 \\
   3 &    1237659324955362294 & 245.899988 &  39.407002 &  19.877 &  1 & $0.51362\pm0.00018$ &  2 &  0 &  0 \\
   4 &    1237659324955361802 & 245.901447 &  39.457362 &  21.091 &  1 & $2.72922\pm0.00132$ &  2 &  0 &  0 \\
   5 &    1237659324955362292 & 245.902653 &  39.409422 &  20.558 &  1 & $0.51274\pm0.00014$ &  2 &  0 &  0 \\
   6 &    1237665355084988891 & 245.904406 &  39.612085 &  20.258 &  1 & $2.42312\pm0.00072$ &  2 &  0 &  0 \\
   7 &    1237659324955427115 & 245.920292 &  39.311703 &  17.286 &  0 & $0.21968\pm0.00001$ &  2 &  0 &  0 \\
   8 &    1237659324955427099 & 245.924866 &  39.333424 &  17.175 &  1 & $0.13229\pm0.00003$ &  2 &  0 &  0 \\
   9 &    1237659324955361592 & 245.929057 &  39.417682 &  16.706 &  0 & $0.10142\pm0.00003$ &  2 &  0 &  0 \\
  10 &    1237659324955427194 & 245.942664 &  39.227732 &  20.642 &  1 & $0.64250\pm0.00023$ &  2 &  0 &  0 \\
\enddata

\tablenotetext{a}{(0) Extended source, (1) Point source.}
\tablenotetext{b}{(1) This study, (2) SDSS DR12, (3) \citet{RG2008}, (4) \citet{Rines_etal2002}, 
(5) \citet{Hill_Oegerle1998}, (6) \citet[][NOAO Fundamental Plane Survey]{Smith_etal2004}, (7) \citet{Zabludoff_etal1990},
(8) \citet[][2MASS Redshift Survey]{Huchra_Etal2012}, (9) \citet{Pustilnik_etal1999}, (10) \citet{Bernardi_etal2002}, (11) \citet{Wyithe_etal2005}.}
\tablenotetext{c}{(0) A2199 non-member, (1) A2199 member.}
\tablenotetext{d}{(0) Not involved with groups, (1-32) Group identification.}
\tablecomments{This table is available in its entirety 
in a machine-readable form in the online journal. 
A portion is shown here for guidance regarding its form and content.}
\label{tab_samp}
\end{deluxetable*}

\section{PHYSICAL PROPERTIES OF CLUSTER GALAXIES IN A2199}
\subsection{Cluster Membership with the Caustic Technique}
The distribution of cluster galaxies in a phase space 
	defined by radial velocity and projected clustercentric radius 
	shows a characteristic trumpet-shaped pattern.
The edges of this distribution are called caustics and are related to
	the escape velocity at each radius \citep{Kaiser1987,RG1989,Diaferio_Geller1997}.
The caustic technique defines the trumpet-shaped pattern, and
	separates galaxies bound in a cluster from foreground and background galaxies \citep{Diaferio_Geller1997,Diaferio1999}.
The technique uses an adaptive kernel to compute a smoothed, two-dimensional density distribution in the phase space,
	and the location of the caustics is where the density reaches a certain threshold.
A detailed explanation on this technique is in \citet{Diaferio1999} and \citet{Serra_etal2011}.
This technique is widely used in identifying cluster galaxies and
	in determining cluster mass profiles \citep{Rines_etal2013,Rines_etal2016,Biviano_etal2013}.
\citet{SD2013} used 100 mock clusters from a cosmological N-body simulation,
	and demonstrated that the caustic technique works well
	in identifying cluster member galaxies
	with the fraction of identified true members to be 95$\pm$3\% within $3\,R_{200}$ \citep[see also][for more details]{Diaferio1999,Serra_etal2011}.
We use a free software tool, The Caustic App v1.2 developed by Anna Laura Serra \& Antonaldo Diaferio,
	to apply this method to the sample of galaxies with measured redshifts at $R<400^\prime$
	and identify the member galaxies of A2199.

\begin{figure*}
\centering
\includegraphics[width=0.78\textwidth]{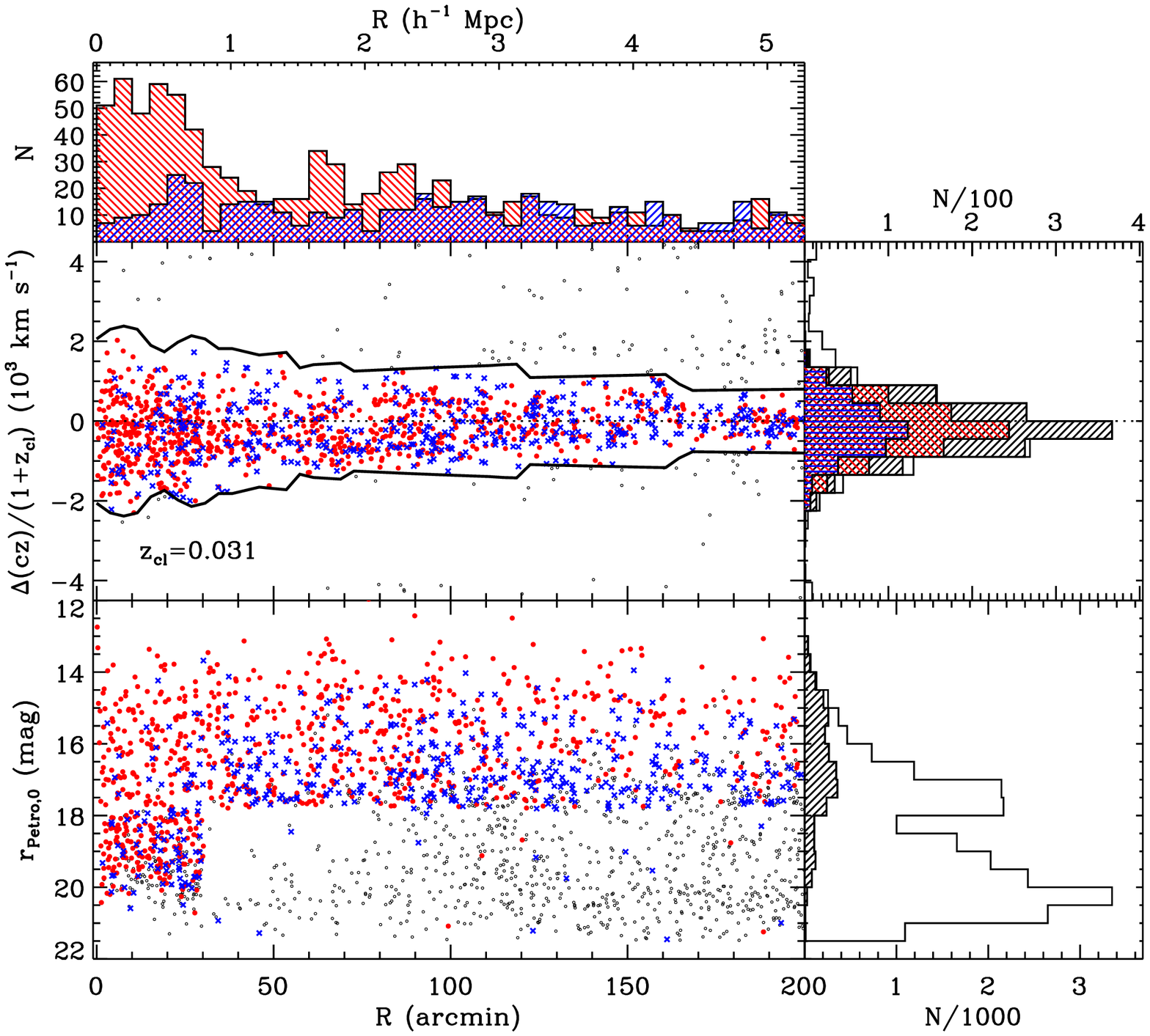}
\caption{(Middle left)  
Rest-frame clustercentric velocities of galaxies  
as a function of projected clustercentric radius.
Black thick lines are the caustics. 
Cluster member galaxies are defined as those within the caustics.
Red dots and blue crosses are the red and blue members, respectively.
Galaxies with measured redshifts that are classified as non-members
are represented with black dots.
Histograms of clustercentric radii (top)
and of rest-frame clustercentric velocities (middle right) 
of galaxies are also shown.
Red and blue hashed histograms are for red and blue members, respectively.
Black hashed histogram is for all members 
and plain one is for galaxies including both members and non-members.
(Bottom left) $r_\textrm{\fns petro,0}$ of galaxies 
as a function of clustercentric radius.
Black dots are non-member galaxies with measured redshifts.
We display only 10\% of them for clarity.
(Bottom right) Histogram of $r_\textrm{\fns petro,0}$.
Black hashed one is for members and plain one is 
for all galaxies with measured redshifts.
}
\label{fig_env}
\end{figure*}

The middle panel of Figure \ref{fig_env} shows the rest-frame clustercentric velocities
	of galaxies as a function of projected clustercentric radius
	with the determined caustics (black solid lines).
The expected trumpet-shaped pattern is obvious, 
	and roughly agrees with the lines based on a visual impression.
The caustic technique also determines a hierarchical center of the cluster
	based on a binary tree analysis \citep[see Appendix A of][for more details]{Diaferio1999}.
The cluster center determined from this technique is R.A.$=247^\circ.135032$ and decl.$=39^\circ.520795$,
	consistent with the X-ray center used in this study.

We finally identify 585 members at $R<60^\prime$ ($R<1.58h^{-1}$Mpc) using the caustic technique.
The red and blue members defined in Figure \ref{fig_ccmr} are plotted, respectively,
	as red circles and blue crosses in the middle panel of Figure \ref{fig_env}.
The bottom panel shows $r$-band apparent magnitudes of galaxies
	as a function of projected clustercentric radius
	(red members: red circles, blue members: blue crosses, non-members with measured redshifts: black circles).
Two different magnitude limits for spectroscopy are apparent at different clustercentric radius ranges:
	$r_\textrm{\fns petro,0}\sim20.5$ at $R<30^\prime$ (MMT/Hectospec in this study) and
	$r_\textrm{\fns petro,0}\sim17.77$ at $30^\prime<R<60^\prime$ (SDSS main galaxy sample).

\subsection{Velocity Dispersion Profile}
\begin{figure*}
\centering
\includegraphics[width=0.7\textwidth]{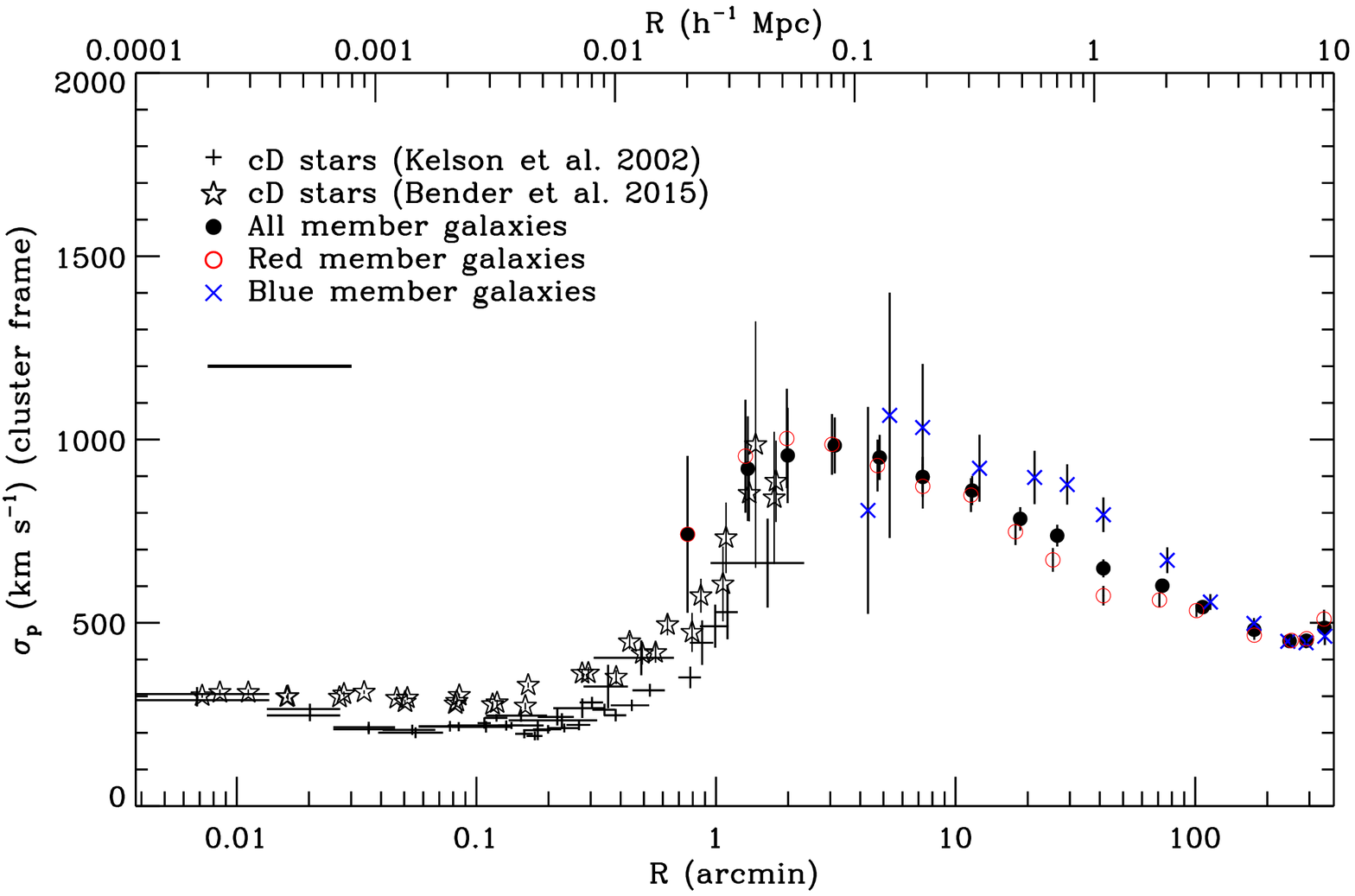}
\caption{Velocity dispersion profile for A2199 as a function of
clustercentric radius.
Points and stars at radii$\lesssim2'$ with $1\sigma$ error bars 
show the velocity dispersion profiles of the cD galaxy 
from \citet[][see their Table 2]{Kelson_etal2002}
and \citet[][see their Table 1]{Bender_etal2015}, respectively.
At larger radii, the velocity dispersion profile for the cluster members
selected by the caustic method is presented.
The horizontal bar shows the width of the overlapping logarithmic bins (0.6 dex).
Black points, red open circles, and blue crosses denote all the members,
the red members, and the blue members, respectively.
The error is estimated by the bootstrap method.
}
\label{fig_disp}
\end{figure*}

The velocity dispersion profile of galaxy clusters provides important information
	on the total mass distribution in clusters or on the orbital velocity anisotropy of cluster galaxies
	\citep{Mahdavi_etal1999,Biviano_Katgert2004,HL2008}.
Moreover, \citet{Geller_etal2014} suggested that the combination of two kinematic tracers
	(e.g. stars of cD galaxy and cluster galaxies) potentially provide 
	complementary measures of the cluster potential \citep[see also][]{Kelson_etal2002}.
We therefore present the velocity dispersion profile of the A2199 galaxies
	with the stellar dispersion profile of the cD galaxy in Figure \ref{fig_disp}.
Red open circles and blue crosses are for the red and blue members, respectively,
	and black filled circles are for all the members.
The points at $R<2^\prime$ are for the cD stars: 
	plus signs from \citet{Kelson_etal2002} and star symbols from \citet{Bender_etal2015}.
The two different velocity dispersion profiles of stars and galaxies
	overlap smoothly and show a turnover around $0.1h^{-1}$Mpc.
This is similar to the results in \citet{Kelson_etal2002} for A2199
	and in \citet{Geller_etal2014} for A383.
\citeauthor{Kelson_etal2002} tried to fit to the observed velocity dispersion profiles
	with one- and two-component mass models, and could obtain an acceptable fit
	when they treat the stellar and dark matter components separately.
A detailed dynamical modeling including the fit 
	to the combined cD and galaxy velocity dispersion profiles
	is beyond the scope of this paper, but we provide the data for the profile of galaxies
	in Table \ref{tab_disp} for future studies.

The velocity dispersion profiles of the total and red samples agree well within the uncertainty.
The blue galaxies seem to have systematically
	higher velocity dispersions than the red galaxies.
However, the inclusion of the blue population to the total sample
	does not make the total velocity dispersion profile significantly different from 
	that of the red sample.
This means that the red members are reliable tracers for the cluster velocity dispersion
	(and cluster mass estimates), which is consistent with the results
	in other studies \citep{Rines_etal2013,Geller_etal2014}.

\subsection{Luminosity Function}
The galaxy luminosity function and its environmental dependence are fundamental tools
	for understanding galaxy formation and evolution \citep{White_Rees1978,Cole1991,White_Frenk1991}.
In particular, a robust measurement of the faint-end slope of the luminosity function
	that provides strong constraints on galaxy formation models is one of key challenges \citep[e.g.][]{Geller_etal2012}.
Typical measurements of the faint-end slope of the luminosity function from observations are
	$-1.6<\alpha<-1.1$ \citep[e.g.][]{Efstathiou_etal1988,Liu_etal2008},
	flatter than the one of the mass function of dark matter subhalos 
	from $\Lambda$CDM simulations ($\alpha=-1.9$, \citealt{Springel_etal2008}).
This discrepancy can be understood by various physical processes relevant to galaxy formation and evolution
	in a dark matter halos including 
	gas cooling, cosmic re-ionization, feedback processes, galaxy merging, and thermal conduction 
	\citep{Benson_etal2003b,Cooray_Milosavljevic2005,Croton_etal2006}.

Interestingly, some studies suggest that the galaxy luminosity function in clusters
	shows an upturn at the faint end (e.g. $-2.1<\alpha<-1.6$, 
	\citealt{Driver_etal1994,dePropris_etal1995,Adami_etal2007,Jenkins_etal2007,Milne_etal2007,
			Yamanoi_etal2007,Banados_etal2010,Agulli_etal2014,Moretti_etal2015,Lan_etal2016}).
There are some physical processes including tidal interactions and shielding from the ultraviolet radiation in clusters
	that could create and protect dwarf galaxies, which result in the excess of dwarf populations;
	this makes the faint-end slope of the luminosity function steep 
	\citep{BH1992,Bekki_etal2001,Tully_etal2002,Benson_etal2003b,Popesso_etal2006}.
\citet{Lan_etal2016} has recently claimed that this faint-end upturn of the luminosity function
	is universal regardless of environment.
However, there are no studies that directly show such an upturn in cluster environment
	using the spectroscopic data \citep[e.g.][]{RG2008,Ferrarese_etal2016},
	which means that the existence and universality of the faint-end upturn feature are still in debate.
Here, we examine the luminosity function of the cluster galaxies of A2199 with our deep and complete spectroscopic data
	focusing on the faint-end slope.

\begin{deluxetable*}{rr@{\hskip4pt}c@{}rcrr@{\hskip4pt}c@{}rcrr@{\hskip4pt}c@{}r}
\tabletypesize{\footnotesize}
\tablewidth{0pc}
\tablecaption{Velocity dispersion of A2199}
\tablehead{
\multicolumn{4}{c}{All members}   & &   \multicolumn{4}{c}{Red members}   & &   \multicolumn{4}{c}{Blue members} \\
\cmidrule{1-4} \cmidrule{6-9} \cmidrule{11-14}
$R$ ($^\prime$) & \multicolumn{3}{c}{$\sigma_p$ (km s$^{-1}$)}   & &
$R$ ($^\prime$) & \multicolumn{3}{c}{$\sigma_p$ (km s$^{-1}$)}   & &
$R$ ($^\prime$) & \multicolumn{3}{c}{$\sigma_p$ (km s$^{-1}$)}
}
\startdata
      0.76 &        742&$\pm$&       214 & & 
      0.76 &        742&$\pm$&       214 & & 
      4.31 &        807&$\pm$&       282
 \\
      1.36 &        920&$\pm$&       143 & & 
      1.33 &        955&$\pm$&       154 & & 
      5.31 &       1066&$\pm$&       334
 \\
      1.99 &        957&$\pm$&       130 & & 
      1.98 &       1003&$\pm$&       135 & & 
      7.29 &       1033&$\pm$&       173
 \\
      3.13 &        984&$\pm$&        76 & & 
      3.05 &        987&$\pm$&        83 & & 
     12.59 &        922&$\pm$&        91
 \\
      4.82 &        951&$\pm$&        62 & & 
      4.72 &        929&$\pm$&        71 & & 
     21.37 &        897&$\pm$&        73
 \\
      7.29 &        898&$\pm$&        55 & & 
      7.30 &        872&$\pm$&        60 & & 
     29.24 &        877&$\pm$&        55
 \\
     11.72 &        861&$\pm$&        39 & & 
     11.59 &        848&$\pm$&        46 & & 
     41.40 &        795&$\pm$&        47
 \\
     18.61 &        784&$\pm$&        32 & & 
     17.79 &        749&$\pm$&        36 & & 
     76.45 &        671&$\pm$&        35
 \\
     26.56 &        738&$\pm$&        30 & & 
     25.47 &        672&$\pm$&        32 & & 
    115.75 &        557&$\pm$&        21
 \\
     41.38 &        649&$\pm$&        25 & & 
     41.38 &        574&$\pm$&        26 & & 
    175.97 &        498&$\pm$&        15
 \\
     72.92 &        601&$\pm$&        17 & & 
     70.94 &        562&$\pm$&        20 & & 
    243.59 &        450&$\pm$&        14
 \\
    107.44 &        543&$\pm$&        13 & & 
    101.29 &        533&$\pm$&        17 & & 
    290.36 &        446&$\pm$&        15
 \\
    176.27 &        481&$\pm$&        10 & & 
    176.50 &        466&$\pm$&        12 & & 
    347.55 &        464&$\pm$&        24
 \\
    247.81 &        450&$\pm$&         9 & & 
    251.90 &        452&$\pm$&        13 & & 
           &           &     &           
 \\
    291.14 &        451&$\pm$&        10 & & 
    291.85 &        457&$\pm$&        13 & & 
           &           &     &           
 \\
    346.34 &        486&$\pm$&        17 & & 
    345.00 &        510&$\pm$&        25 & & 
           &           &     &           
 \\
\enddata

\tablecomments{Errors are estimated by the bootstrap method.}
\label{tab_disp}
\end{deluxetable*}

\citet{RG2008} used the redshift data of the central region of A2199 ($R<30^\prime$) 
	from MMT/Hectospec observations to derive the luminosity function at $r<20$ mag.
We use the redshift data from our observation that is deeper and more complete than previous surveys 
	to re-determine the luminosity function of A2199 focusing on the faint-end slope.
Following \citet{RG2008}, we first select the cluster galaxies by simply applying their velocity cut
	(i.e. 7000$<cz<$11000 km s$^{-1}$); 
	there are 400 galaxies by excluding the cD galaxy from the luminosity function analysis.
We count galaxies at each absolute magnitude bin, and correct the counts for spectroscopic incompleteness 
	by weighting each galaxy with the inverse of the completeness in Figure \ref{fig_scompl_r}
	and for surface brightness incompleteness in the SDSS photometric catalog
	by multiplying the correction factor in Fig. 6 of \citet{Blanton_etal2005}.
We fit to the data where the total (surface brightness and spectroscopic) completeness is larger than 50\%
	(i.e. at $r_\textrm{\fns petro,0}<20.3$)
	with the Schechter function \citep{Schechter1976}:
\begin{equation}
   \phi(M)dM = \phi^\ast 10^{0.4(1+\alpha)(M^\ast-M)}\textrm{exp}\left(-10^{0.4(M^\ast-M)}\right)dM
\label{eqn_LF}
\end{equation}
	where $M$ is an absolute magnitude, 
	$M^\ast$ is the characteristic magnitude of the luminosity function,
	and $\alpha$ is the faint-end slope.
We use the MPFIT package in IDL \citep{Markwardt2009} 
	to determine the best-fit Schechter function, 
	and compute the uncertainties of the best-fit parameters 
	by repeating the fitting procedure 1000 times with re-sampled data sets.
The resulting best-fit parameters
	are $M^\ast\footnote{
		All absolute magnitudes in this paper are based on $H_0=70\, \textrm{km s}^{-1} \textrm{Mpc}^{-1}$.} 
	=-21.68\pm0.64$
	and $\alpha=-1.26\pm0.06$.
This faint-end slope is slightly steeper than that the one in \citeauthor{RG2008} (2008, $\alpha=-1.13^{+0.07}_{-0.06}$),
	but is much shallower than the one expected from the faint-end upturn of the luminosity function (i.e. $-2.1<\alpha<-1.6$).

On the other hand, the velocity cut used above corresponds to
	$-2298 < \Delta(cz)/(1+z_\textrm{\fns cl}) < 1581\,\textrm{km s}^{-1}$;
	this criterion is different from the caustics defined in Figure \ref{fig_env}.
To examine how this selection changes the result,
	we also derive the luminosity function using the member galaxies
	identified with the caustic technique (i.e. 386 members at $R<30^\prime$).
The best-fit parameters are 
	$M^\ast=-21.68\pm0.80$ and $\alpha=-1.26\pm0.06$,
	which shows no significant differences from the case based on the radial velocity cut.

\begin{figure*}
\centering
\includegraphics[width=0.8\textwidth]{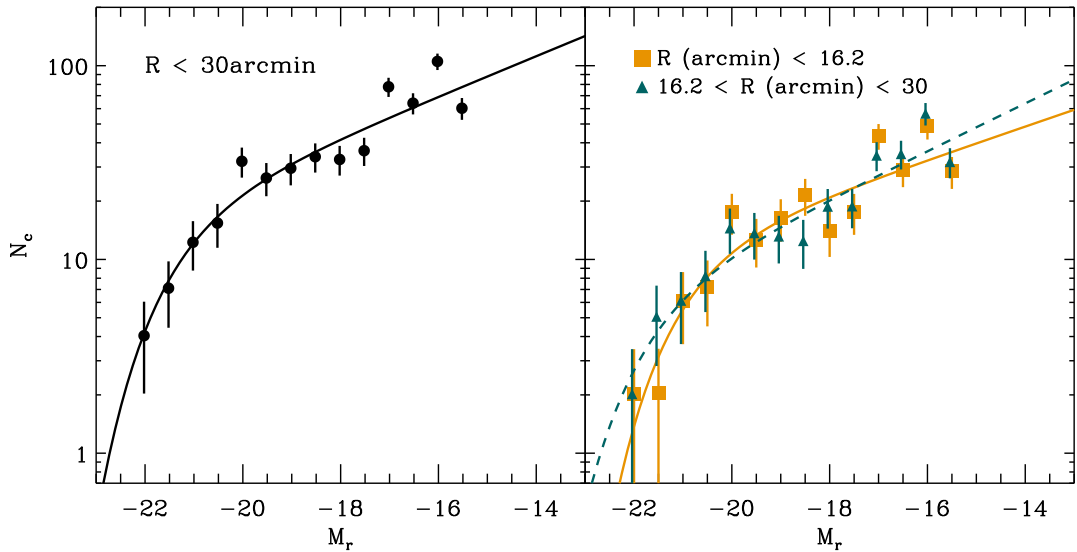}
\caption{Luminosity functions of A2199 member galaxies 
corrected for surface brightness and spectroscopic incompletenesses.
The cD galaxy is excluded from the analysis,
and the $r$-band absolute magnitude, $M_r$ is based on $H_0=70\,\textrm{km s}^{-1} \textrm{Mpc}^{-1}$.
(Left) Luminosity function of galaxies selected by the caustic method
at $R<30^\prime$ is measured
and fitted with a single Schechter function.
Black solid line represents the best-fit result.
Error bars are Poissonian uncertainties.
Data with total (surface brightness and spectroscopic) completeness 
greater than 50\% are plotted and used for the fit,
(Right) Luminosity functions of two subsamples of the member galaxies
(selected by the caustic method as well).
The subsamples are divided by the clustercentric radius.
Yellow squares and green triangles are luminosity functions in the inner and outer regions, respectively.
Solid and dashed lines represent the best-fit results.
Error bars are Poissonian uncertainties.
}
\label{fig_LF}
\end{figure*}

To examine whether the luminosity function changes with environment,
	we show the luminosity functions for inner (yellow squares) and outer (green triangles) regions
	in the right panel of Figure \ref{fig_LF}.
The radial ranges are chosen to have similar numbers of galaxies in the regions.
We also fit to the data with the high spectroscopic completeness at $M_r<-15$ (i.e. filled symbols),
	and obtain the best-fit parameters of 
	$M^\ast=-21.26\pm1.11$ and $\alpha=-1.22\pm0.10$ for the inner region
	and $M^\ast=-22.11\pm3.94$ and $\alpha=-1.31\pm0.13$ for the outer region.
The slopes of the two subsamples agree within the uncertainty,
	indicating no significant changes of the luminosity function 
	with clustercentric radius at $R<30^\prime$.
This is again consistent with the result of \citet{RG2008}.

We found no evidence for the faint-end upturn of the luminosity function in A2199 at $M_r<-15$.
This could be because our magnitude limit is not faint enough to probe the magnitude range
	where the upturn is expected to appear (e.g. $M_r\gtrsim-15$, \citealt{Lan_etal2016}).
However, a recent work by \citet{Ferrarese_etal2016} who used the very deep spectroscopic data
	of the Virgo cluster with $M_g<-9.13$ also found no faint-end upturn in their luminosity function.
It should be noted that many studies suggesting the faint-end upturn of the luminosity function
	are based on photometric data, which can suffer from systematic uncertainties
	including the subtraction of background galaxies, especially in the faint end of the luminosity function.
We refer the readers to \citet{RG2008} for more discussion on the possible uncertainties in determining the luminosity function.
Deeper spectroscopic surveys of galaxy clusters will be helpful for better constraining the faint-end slope of the luminosity function.

\section{COMPARISON BETWEEN GALAXIES AND THE INTRACLUSTER MEDIUM}
\subsection{Velocity Structure in A2199}
The kinematics of cluster galaxies can provide an important hint of merging history of galaxy clusters \citep{HL2009}.
Moreover, because of different dynamical properties of galaxies and the ICM 
	(i.e. collisionless galaxies and the collisional ICM),
	comparison of the bulk motion between the two is also useful for 
	understanding the dynamical state of galaxy clusters \citep[e.g.][]{Roettiger_Flores2000}.

\begin{figure*}
\centering
\includegraphics[width=0.75\textwidth]{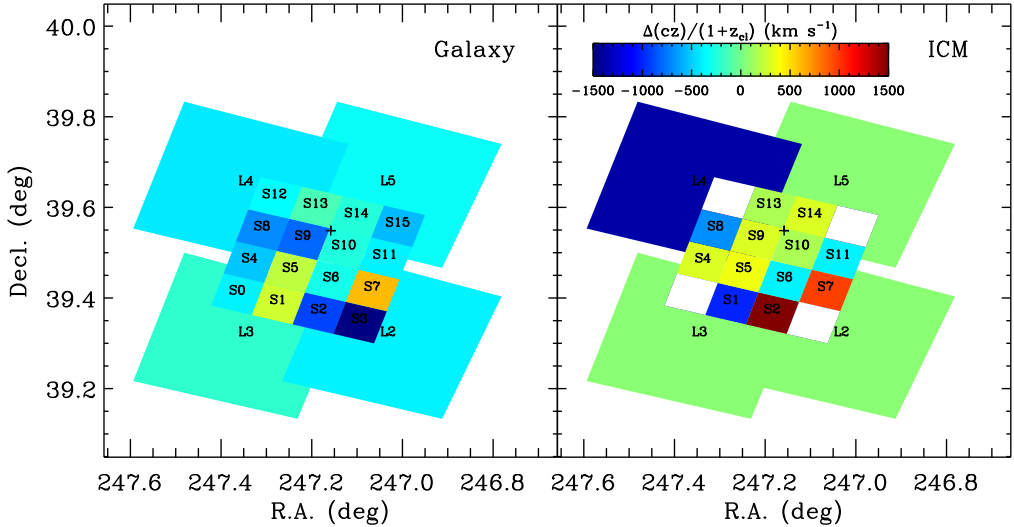}
\caption{Two-dimensional maps color-coded by radial velocities 
of galaxies (left) and the ICM (right) in the cluster rest frame
as a function of right ascension and of declination.
The radial velocities of galaxies and the ICM are measured within each cell.
Empty cells indicate the regions with no measurements.
For galaxies, those with $|\Delta cz|/(1+z_\textrm{\fns cl})<3000\,\textrm{km s}^{-1}$ are used.
The mean value of the radial velocities of the galaxies
in each region is calculated.
Plus symbols refer to the cluster center.
}
\label{fig_velmap}
\end{figure*}

\begin{figure*}
\centering
\includegraphics[width=0.75\textwidth]{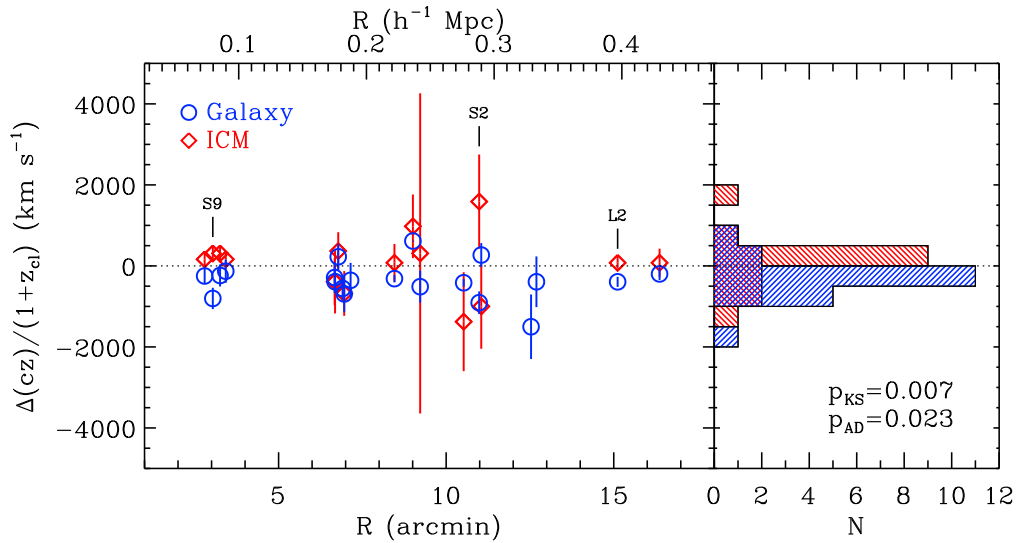}
\caption{Radial velocities
of galaxies (blue) and the ICM (red) in the cluster rest frame
measured in the regions of Figure \ref{fig_velmap},
as a function of clustercentric radius (left),
and their histogram (right).
Error bar for the ICM is an $1\sigma$ statistical error 
in the measurement of radial velocity, and that for galaxies is
the standard error of the mean.
}
\label{fig_vel_r}
\end{figure*}

To compare the kinematics of galaxies with that of the ICM in A2199,
	we use the radial velocity measurements of the ICM in \citet{OY2016}.
They measured the radial velocities of the ICM in different regions of A2199
	from the analysis of emission lines in the X-ray spectra taken with the Suzaku satellite.
To compare the radial velocities of the cluster galaxies with those of the ICM,
	we determine the average velocity of the galaxies in each region
	where the ICM radial velocity is measured;
	the average radial velocity of the galaxies is the mean
	of the velocity distribution in each cell 
	(see Appendix for the radial velocity distribution of galaxies in each cell),
	and that of ICM is the radial velocity determined from the stacked X-ray spectrum in each cell.
Figure \ref{fig_velmap} shows the positions of such regions (left: galaxies, right: the ICM)
	delineated by different sized squares: 
	four large cells of $18^\prime\times18^\prime$ FOV 
	and sixteen small cells of $4^\prime.5\times4^\prime.5$ FOV.
Each cell is color-coded by the average radial velocities 
	of the galaxies and the ICM in the cluster rest frame.
There are four empty small cells for the ICM without velocity measurements
	because of large uncertainties near X-ray CCD edges.
We note that the radial velocity of the ICM in the upper left large cell, L1, is
	very different from the other three large cells ($\Delta v\sim1500\textrm{km s}^{-1}$).
However, the velocity measurement error in this cell is also large ($\sigma_v\sim1300\textrm{km s}^{-1}$),
	indicating the large velocity offset is not statistically significant 
	(see Figure 1 in \citealt{OY2016} for more details).
There are some cells where the galaxies and the ICM
	move in opposite directions (e.g. S2) as well as those
	where the two components move in the same direction
	but at different speeds.

To compare the radial motions of the two components quantitatively, 
	we plot the radial velocities of the galaxies and the ICM in the cells
	as a function of projected clustercentric radius.
In the left panel of Figure \ref{fig_vel_r}, 
	we show the radial velocities of the galaxies and the ICM for each cell
	with blue circles and red diamonds, respectively.
The error bar for the ICM is an 1$\sigma$ statistical error 
	in the measurement of radial velocity,
	and that for the galaxies is the standard error of the mean.
There are three cells where the difference in the radial velocity
	between the galaxies and the ICM is significant more than ${\sim}2\sigma$:
	S9($4.3\sigma$), S2($2.0\sigma$), and L2($3.2\sigma$).
We mark these cells with vertical lines and their names in Figure \ref{fig_vel_r}.
We perform the Kolmogorov-Smirnov (K-S) two sample test 
	and the Anderson-Darling k-sample test (A-D)
	for the distributions of the radial velocities of the galaxies and the ICM 
	to determine whether these two distributions are drawn 
	from the same distribution (null hypothesis).
The $p$-values from the two tests are 0.007 and 0.023, respectively,
	suggesting that there is a possible difference in the radial velocity
	between the galaxies and the ICM with $2.0{-}2.5\sigma$ significance levels.
This possible difference between the two components could be related 
	to the recent merging activity of A2199 that is evidenced by gas sloshing
	probably caused by a minor merger \citep{Nulsen_etal2013}.
Because the possible systematic difference in the velocity measurements of the galaxies and the ICM
	is not negligible, future X-ray observations with better velocity measurement capability
	will be useful for drawing a strong conclusion \citep[e.g.][]{Kitayama_etal2014}.

\begin{figure*}
\centering
\includegraphics[width=0.75\textwidth]{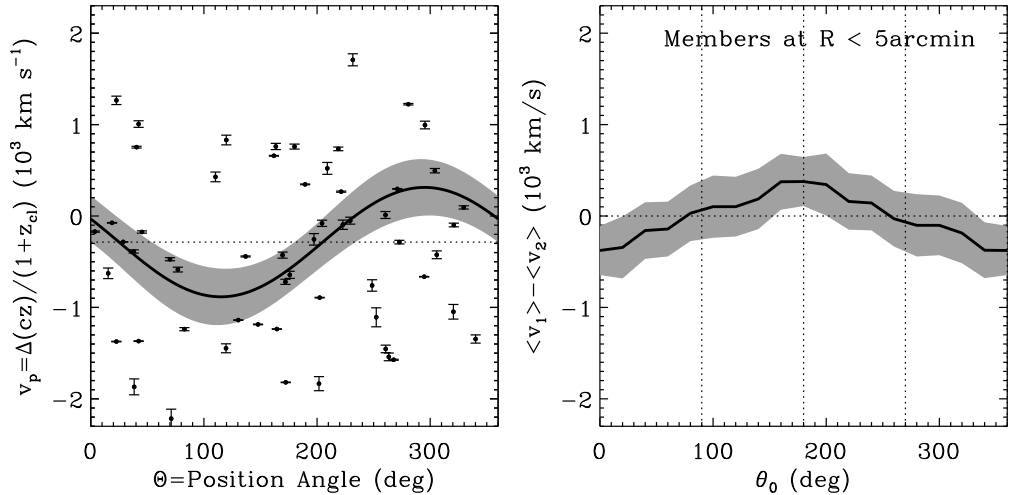}
\caption{
(Left) Radial velocities of member galaxies at $R<5^\prime$ (filled circles with error bar) 
in the cluster rest frame as a function of (projected) position angle.
Black solid line indicates the best-fit rotation curve with Equation (\ref{eqn_rot})
and gray band represents 1$\sigma$ range.
(Right) Difference between two median radial velocities ($<v_1>$ and $<v_2>$) 
of two galaxy subsamples divided by an axis with position angle $\theta_0$. 
The velocity difference reaches its maximum ($377\pm268\, \textrm{km s}^{-1}$) at $\theta_0{\sim}180^\circ$.
}
\label{fig_rot}
\end{figure*}

One interesting feature is that the average radial motion of the galaxies in S9 cell
	is discrepant when compared to the surrounding cells (i.e. S10, S13, S14),
	whereas the average radial motion of the ICM shows no such discrepancy.
This radial motion of the galaxies could result from
	bulk motion including the global rotation of clusters.
We therefore quantify the rotation of the galaxies at $R<5^\prime$ around the center of A2199
	by fitting the radial velocities of the galaxies ($v_p=\Delta(cz)/(1+z_\textrm{\fns cl})$)
	with a function of position angle ($\Theta$),
\begin{equation}
v_p(\Theta) = v_\textrm{\fns sys} + v_\textrm{\fns rot}\,\textrm{sin}(\Theta-\Theta_0),
\label{eqn_rot}
\end{equation}
	following \citet{HL2007}.
$v_\textrm{\fns rot}$ and $\Theta_0$ are fitting parameters,
	which represent the (projected) rotational speed 
	and the position angle of the rotation axis, respectively.
We fix $v_\textrm{\fns sys}$ with the median value of the cluster galaxies at $R<5^\prime$ (dotted line in the figure).
The left panel of Figure \ref{fig_rot} shows the radial velocities of the galaxies at $R<5^\prime$
	as a function of position angle with the best-fit rotation curve (thick solid line).
The best-fit parameters are $v_\textrm{\fns rot}=598\pm264\,\textrm{km s}^{-1}$ and $\Theta_0=205\pm36\,^\circ$,
	suggesting a possible rotational signal with $2.3\sigma$.

As a sanity check, 
	we use another method to detect any rotation signal of galaxy clusters \citep{Manolopoulou_Plionis2016}.
We examine the difference between the median velocities ($<v_1>$ and $<v_2>$) 
	of two galaxy subsamples divided by an axis that passes through the cluster center.
The right panel of Figure \ref{fig_rot} shows the velocity difference 
	as the division axis rotates consecutively (rotation diagram).
$\theta_0$ is the position angle of the axis.
If a cluster is rotating, the rotation diagram should show a clear periodic trend,
	having its maximum or minimum when the division axis coincides with the rotation axis.
In general, the maximum corresponds to the rotational speed \citep[see Section 2.1 of][]{Manolopoulou_Plionis2016}.
The right panel of Figure \ref{fig_rot} shows a periodic change of the velocity difference ($<v_1>-<v_2>$)
with $\theta_0$ with the maximum of $390\pm268\,\textrm{km s}^{-1}$ at $\theta_0\sim180^\circ$.

\begin{figure*}
\centering
\includegraphics[width=0.9\textwidth]{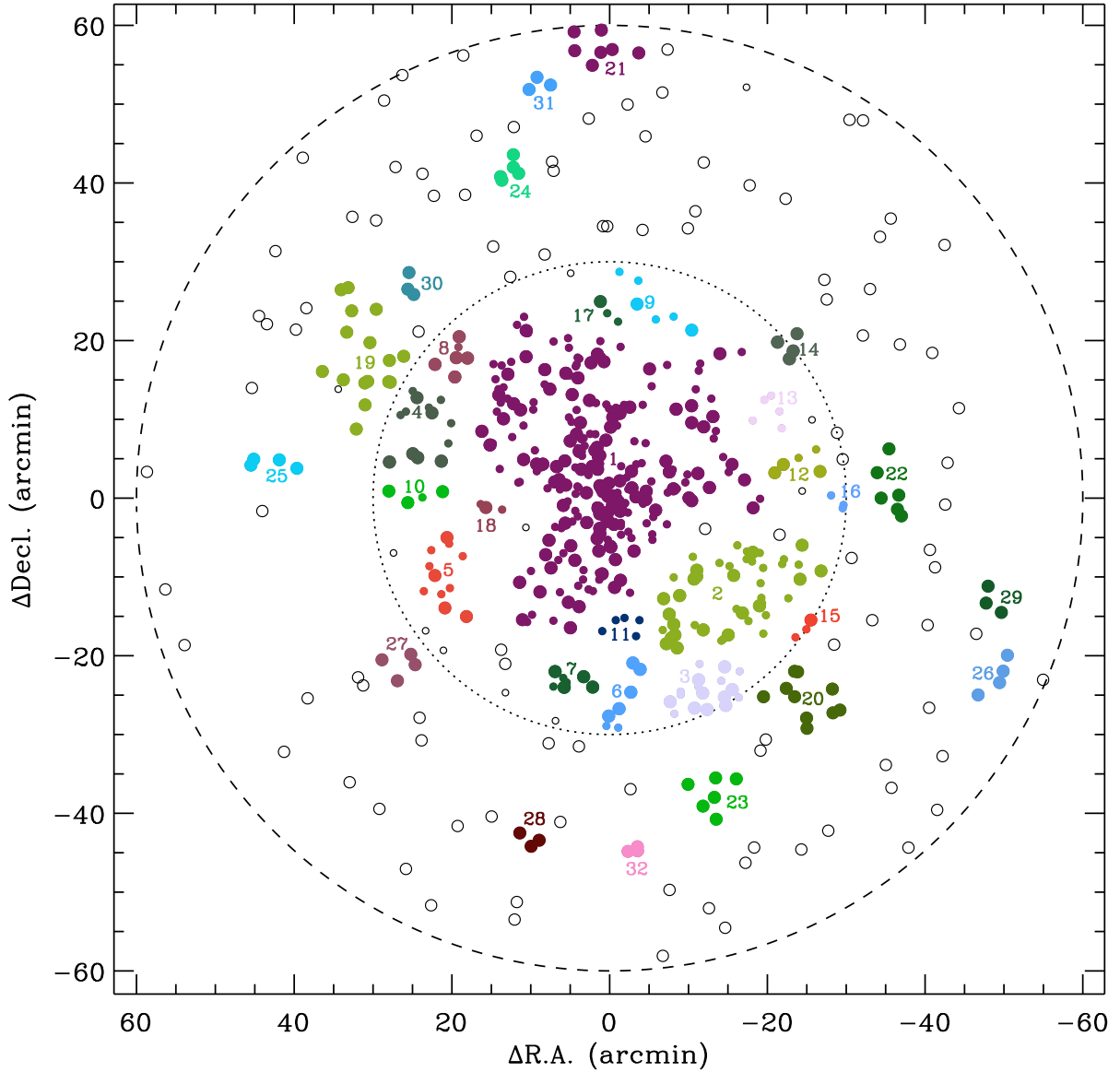}
\caption{Spatial distribution of friends-of-friends groups in and around A2199.
Filled and open circles denote galaxies
with $|\Delta cz|/(1+z_\textrm{\fns cl})<3000\textrm{km s}^{-1}$.
Galaxies fainter than $r_\textrm{\fns Petro,0}=17.77$ are denoted by smaller circles,
and galaxies brighter than $r_\textrm{\fns Petro,0}=17.77$ are denoted by bigger circles.
Big circles with dotted and dashed lines represent 
clustercentric radii of $30^\prime$ and $1^\circ$, respectively.
x-axis and y-axis are angular distances from the X-ray peak position 
(R.A.=247.135041$^\circ$ and decl.=39.520810$^\circ$)
along right ascension and declination directions, respectively.
At $R=30^\prime$, limiting apparent magnitude ($r_\textrm{\fns petro,0}$) of the data changes
from 20.5 (inner region) to 17.77 (outer region) 
(see bottom left panel of Figure \ref{fig_env}).
Filled circles are members of subgroups
and open circles are non-members.
Each subgroup is distinguished by color.
Subgroup ID in Table \ref{tab_grp} is also shown.
}
\label{fig_fofgrp}
\end{figure*}

The rotation of the cluster galaxies at $R<5^\prime$ is 
	detected at a significance level of 2.3$\sigma$ by the method of \citet{HL2007}
	and of 1.5$\sigma$ by the method of \citet{Manolopoulou_Plionis2016}.
The measurements of the rotational amplitude and the rotation axis by the two methods 
	agree within the uncertainty.
As expected, the rotation axis we find is placed 
	to separate the S9 cell from the other cells at similar radii (S10, S13 and  S14) in Figure \ref{fig_velmap}.
It is interesting to note that only the galaxies show a hint of rotation in A2199 unlike the ICM.
\citet{Poole_etal2006} suggested that the kinematics of the ICM can be affected by the disruptive gas dynamical forces
	that do not act on the dynamically dominant component (i.e. dark matter, galaxies),
	which makes the ICM remove the signature of substructures faster than the other components.
On the other hand, a numerical simulation of cluster major mergers 
	(e.g. mass ratio of $\sim2.5:1$, \citealt{Roettiger_Flores2000})
	shows that the rotation of the ICM survives longer than that of the galaxies
	when a cluster has experienced an off-axis major merging.
Another numerical simulation also shows that the rotation of the ICM in a relaxed cluster
	is stronger than that of galaxies at all radii \citep{Baldi_etal2017}.
These results from numerical simulations appear to differ from the case of A2199 in this study,
	suggesting that the rotation of the galaxies in the central region of A2199 (i.e. $R<5^\prime$)
	is not induced by an off-axis major merger.
Instead, a recent minor merger could be responsible for the different radial motions 
	between the galaxies and the ICM in A2199.
Indeed, \citet{Nulsen_etal2013} found evidence of a recent minor merger event (400 Myr ago)
	in the central region of A2199 with Chandra observations,
	supporting our conclusion.
Because both the galaxy number density and the X-ray intensity maps 
	show no distinctive clumps in the central region (see Figure \ref{fig_galden_xray_center}),
	the minor merger might result from an infall of a disrupted galaxy group;
	this could not affect the kinematics of the ICM,
	but could affect only the kinematics of the galaxies.
In addition, AGN in cD galaxies can give a significant impact 
	on the inner ICM and its dynamical state \citep{Cui_etal2016,Cui_etal2017}.
For example, the AGN feedback including jets can transfer the extra energy to the ICM,
	which can make the velocity field of the inner ICM sufficiently disordered
	\citep[e.g.][]{Heinz_etal2010,Baldi_etal2017}.
Indeed, \citet{Nulsen_etal2013} showed that there are complex interactions between
	radio outbursts from the AGN in the cD galaxy of A2199 and the inner ICM,
	supporting this argument.

\subsection{Galaxy Groups in the Central Region of A2199}
In this section, we identify galaxy groups in the central region of A2199
	using the galaxy catalog in this study, and compare their spatial distribution with that of the ICM.
Identifying substructures (i.e. subhalos) in galaxy clusters and characterizing their physical properties 
	are important to understand the mass assembly history of clusters \citep{Okabe_etal2014}.
The comparison of spatial distributions between optically-selected galaxy groups with X-ray clumps (and with weak-lensing peaks)
	is an important step to understand the reliability of the group identification method 
	and the systematics of each method.
To do that, we use Suzaku X-ray images in the 0.7--2.0 keV band of A2199 (T. Tamura et al. 2017, in prep.).
However, the incomplete spatial coverage of the Suzaku observations (see Figure \ref{fig_fofgrp_xray})
	prevents us from a detailed, quantitative comparison between the galaxies and the ICM
	(e.g. cross-correlation of the galaxy number map with the X-ray intensity map).
Therefore, we focus on a simple visual identification of X-ray counterparts of optically-selected galaxy groups in this study
	and simple correlation tests between group galaxies and X-ray intensity in clusters.

\subsubsection{Identification of Galaxy Groups with a Friends-of-Friends Algorithm}
\begin{deluxetable*}{rrccrr}
\tabletypesize{\footnotesize}
\tablewidth{0pc} 
\tablecaption{Galaxy group candidates within $1^{\circ}$ from the center of A2199}
\tablehead{
\multirow{2}{*}{Group ID}  & \multirow{2}{*}{N$_\textrm{mem}$\tablenotemark{a}} & R.A.$_\textrm{cen}$ & Decl.$_\textrm{cen}$ & \multicolumn{1}{c}{$cz_\textrm{cen}$} & \multicolumn{1}{c}{$\sigma_{p}$} \\ & & ($^{\circ}$) & ($^{\circ}$) & \multicolumn{1}{c}{(km s$^{-1}$)} & \multicolumn{1}{c}{(km s$^{-1}$)}
}
\startdata
   1 & $   244$ & 247.190208 &  39.603261 &       9121 &        932 \\
   2 & $ \ge47$ & 246.819011 &  39.342433 &       8977 &        765 \\
   3 & $ \ge17$ & 246.904735 &  39.136092 &       8986 &        809 \\
   4 & $ \ge13$ & 247.684846 &  39.723306 &       9101 &        610 \\
   5 & $    11$ & 247.608242 &  39.384192 &       9259 &        382 \\
   6 & $  \ge7$ & 247.132763 &  39.103324 &       9310 &        888 \\
   7 & $     7$ & 247.282277 &  39.158823 &       9080 &        709 \\
   8 & $  \ge6$ & 247.580229 &  39.845116 &       8890 &        871 \\
   9 & $  \ge6$ & 247.078795 &  39.959155 &       9584 &        954 \\
  10 & $  \ge5$ & 247.671460 &  39.557037 &       8570 &        619 \\
  11 & $     5$ & 247.117738 &  39.290438 &       9617 &        615 \\
  12 & $  \ge5$ & 246.639659 &  39.618715 &       9431 &        906 \\
  13 & $     5$ & 246.713557 &  39.731108 &       9706 &        667 \\
  14 & $  \ge5$ & 246.662488 &  39.858683 &       9634 &        233 \\
  15 & $  \ge3$ & 246.621210 &  39.269845 &      10052 &        516 \\
  16 & $  \ge3$ & 246.519715 &  39.532315 &       9358 &        332 \\
  17 & $     3$ & 247.164662 &  39.939751 &       8609 &        247 \\
  18 & $     3$ & 247.497326 &  39.528499 &       8871 &        559 \\
\hline
  19 & $    16$ & 247.831534 &  39.838499 &       8858 &        360 \\
  20 & $    10$ & 246.645921 &  39.127725 &       9349 &        676 \\
  21 & $  \ge7$ & 247.182264 &  40.495055 &       8847 &        404 \\
  22 & $     6$ & 246.391443 &  39.552047 &       9299 &        248 \\
  23 & $     6$ & 246.874044 &  38.943058 &       9657 &        738 \\
  24 & $     5$ & 247.424576 &  40.235407 &       9133 &        551 \\
  25 & $     4$ & 248.134680 &  39.626435 &       8477 &        217 \\
  26 & $     4$ & 246.095654 &  39.177529 &       9285 &        734 \\
  27 & $     4$ & 247.736806 &  39.205308 &       9149 &        793 \\
  28 & $     3$ & 247.371478 &  38.825007 &      10180 &        695 \\
  29 & $     3$ & 246.123905 &  39.321854 &       9142 &        213 \\
  30 & $     3$ & 247.711780 &  39.989630 &       8383 &        908 \\
  31 & $     3$ & 247.359425 &  40.422342 &       8476 &        260 \\
  32 & $     3$ & 247.083254 &  38.802598 &       9270 &         87 \\
\enddata

\tablecomments{Because number density of galaxy sample is different in inner and outer regions
due to the different limiting apparent magnitudes of the data, 
linking length ($ll$) for the friends-of-friends algorithm is chosen differently. 
Group 1-17 are those found at $R<30^\prime$ and identified with $ll=0.083h^{-1}$Mpc, 
and Group 18-32 are those at $R=30{-}60^\prime$ and identified with $ll=0.090h^{-1}$Mpc.}
\tablenotetext{a}{Number of member galaxies in each group. 
For groups close to the boundaries
where the limiting apparent magnitude changes ($R=30^\prime$) and 
the outermost boundary ($R=1^\circ$),
$N_\textrm{\fns mem}$ is underestimated. 
Thus we add $\ge$ to indicate that 
the given $N_\textrm{\fns mem}$ is the lower limit.}
\label{tab_grp}
\end{deluxetable*}

We first identify group candidates in the cluster by applying a friends-of-friends (FoF) algorithm \citep{Huchra_Geller1982} 
	to the sample of the A2199 galaxies at $R<60^\prime$.
Because the radial velocities of cluster galaxies are strongly affected by peculiar velocities,
	we use only R.A. and declination information to connect galaxies 
	after applying a velocity cut to select cluster galaxies.
There are two magnitude limits for the galaxy sample at $R<60^\prime$
	($r_\textrm{\fns petro,0}\sim20.5$ at $R<30^\prime$ from this survey
	and $r_\textrm{\fns petro,0}\sim17.77$ at $R=30{-}60^\prime$ from the SDSS).
Because the galaxy number density differs with the magnitude limit,
	we use two different linking lengths for the samples of the two different magnitude limits
	to identify galaxy groups with similar physical properties.
To determine an optimal linking length for galaxy groups,
	we first use the galaxies with $r_\textrm{\fns petro,0}\le17.77$
	(the limiting apparent magnitude in the outer region at $R>30^\prime$)
	and $|\Delta cz|/(1+z_\textrm{\fns cl})\le3000\,\textrm{km s}^{-1}$
	in a region of A2199 supercluster ($12^\circ\times12^\circ$)
	where previously known galaxy groups and clusters including 
	Abell 2197W/E, NRGs385, NRGs388, NRGs396, and NGC6159 exist.
We then use several linking lengths and choose the one that
	best reproduces the physical properties 
	of the known clusters and groups \citep[see Table 3 of][]{LeeGH_etal2015},
	which is $0.090h^{-1}$Mpc ($ll_1$).
For the inner region at $R<30^\prime$ with a denser galaxy sample,
	we use the galaxies at $r_\textrm{\fns petro,0}<20.5$ with the same velocity cut. 
We again test several linking lengths and choose the one 
	that connects galaxies to form groups with similar spatial extents
	to the groups previously found with $ll_1$,
	which is $ll_2=0.083h^{-1}$Mpc.

\begin{figure*}
\centering
\includegraphics[width=0.9\textwidth]{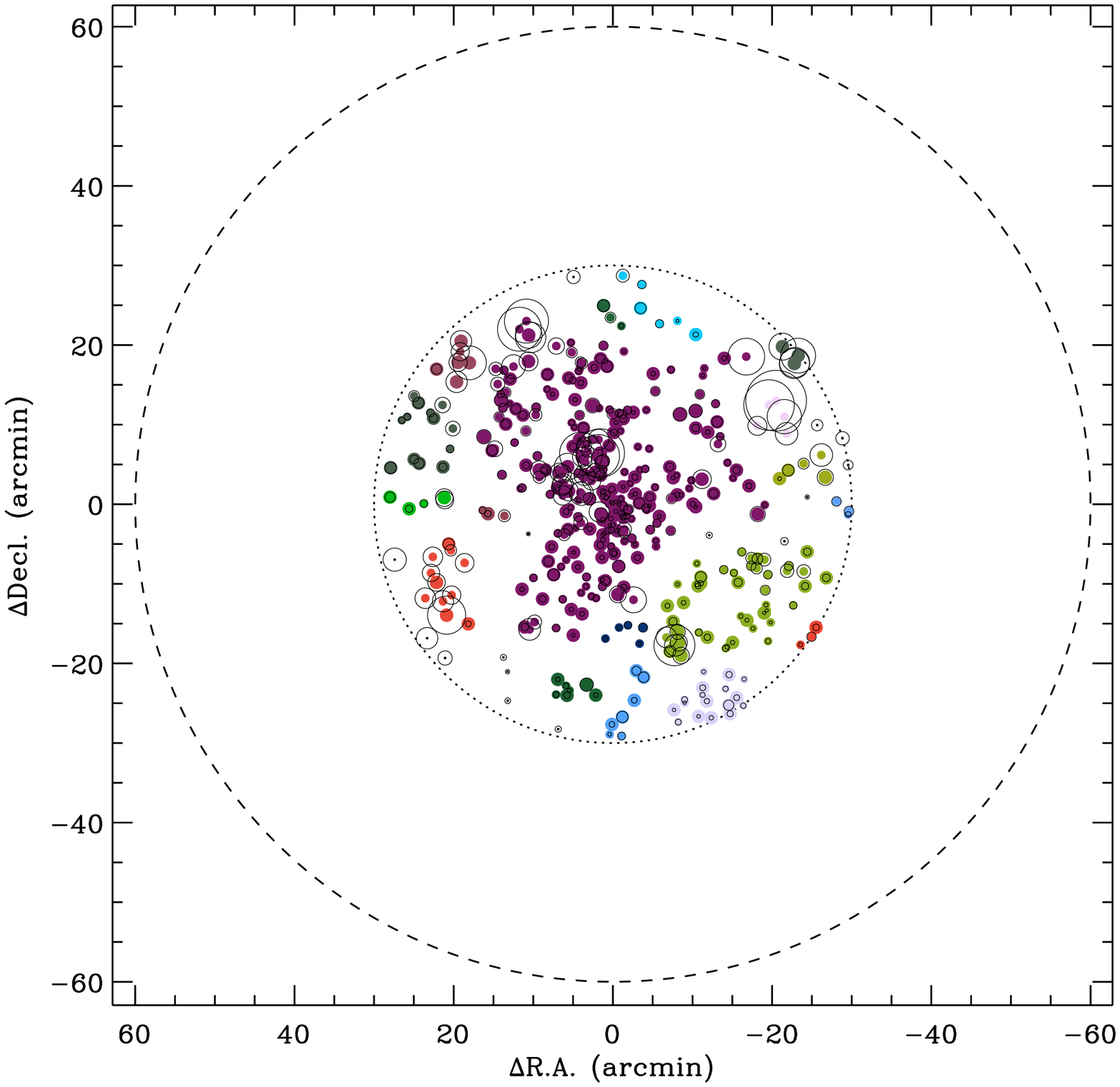}
\caption{Similar to Figure \ref{fig_fofgrp},
but only for the groups at $R<30^\prime$.
Black empty circle refers to kinematic information around each galaxy.
Its size is proportional to $e^\delta$
where $\delta$ denotes how much the kinematics 
of the local environment of a galaxy, defined by 11 nearest neighbors,
is different from the global cluster kinematics (Equation (\ref{eqn_delta})).
}
\label{fig_dsdiagram}
\end{figure*}

Figure \ref{fig_fofgrp} shows the spatial distribution of the A2199 galaxies
	with the identified subgroups that contain at least three members.
The groups at $R<30^\prime$ are from the cluster galaxy sample 
	of $r_\textrm{\fns petro,0}<20.5$ with $ll_2=0.083h^{-1}$Mpc, 
	and those at $R>30^\prime$ are from the sample 
	of $r_\textrm{\fns petro,0}<17.77$ with $ll_1=0.09h^{-1}$Mpc.
When groups identified with $ll_1=0.09h^{-1}$Mpc are distributed 
	across the boundary of $R=30^\prime$,
	we combine them with the neighboring groups at $R<30^\prime$.
Filled circles are members of the subgroups and open circles are non-members.
Bright galaxies with $r_\textrm{\fns petro,0}<17.77$ are indicated with big circles,
	while faint galaxies with $r_\textrm{\fns petro,0}>17.77$ are denoted by small circles.
The groups are distinguished by color and numbered by group id 
	as listed in Table \ref{tab_grp}.
In Table \ref{tab_grp}, we list group id, number of members ($N_\textrm{\fns mem}$),
	central (median) position ($\textrm{R.A.}_\textrm{\fns cen}$, $\textrm{Decl.}_\textrm{\fns cen}$),
	median line-of-sight velocity ($cz_\textrm{\fns cen}$), 
	and line-of-sight velocity dispersion ($\sigma_p$).

Because the FoF groups are found in the projected two-dimensional space,
	there could be some false groups that are not physically associated,
	but look clustered only in the plane of R.A. and declination.
To determine the fraction of false groups in our group catalog, 
	we perform a simple test of comparing FoF groups found in a projected two-dimensional space
	with those found in the real three-dimensional space.
We first construct a set of particles that follows a Navarro-Frenk-White profile \citep[NFW profile;][]{nfw1996} 
	at $R<30^\prime$ with the concentration parameter of A2199 \citep[c=8,][]{Rines_etal2002}.
We keep the total number of the particles to be the same 
	as the total number of the cluster galaxies at $R<30^\prime$ ($N_\textrm{\fns gal}=410$).
We then apply the three-dimensional FoF algorithm to this sample using the linking length $0.083h^{-1}$Mpc.
We then define a false detection rate of group members as
	the fraction of members that belong to any FoF groups 
	found in a projected two-dimensional space 
	but not in FoF groups found in the three-dimensional space.
The test with 1000 data sets results in the false detection rate of $49\%$ 
	with a standard deviation of $3\%$, suggesting a significant contamination.
However, it should be noted that the false detection rate could be lower than this	
	because galaxies in real clusters follow the NFW density profile with an additional subclustering.
This subclustering will reduce the number of galaxies that are not associated with subgroups in clusters.
This effect is not considered in our experiment, which can result in a higher false detection rate than the true value.
We therefore do not claim that our group catalog is clean and complete,
	and do restrict our analysis to simple statistical tests.

\begin{figure*}
\centering
\includegraphics[width=0.83\textwidth]{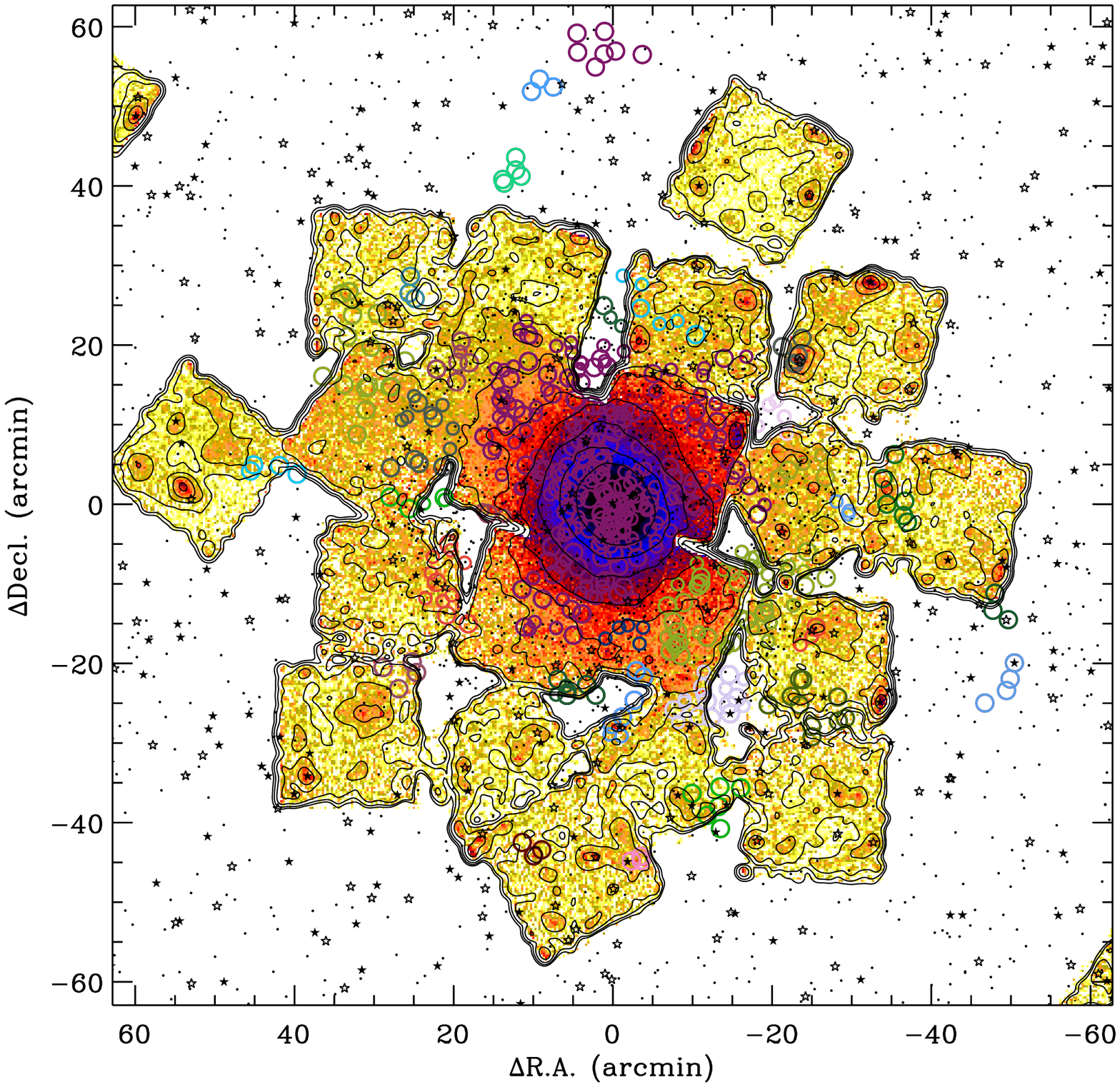}
\caption{Spatial distribution of the FoF groups 
on top of the Suzaku X-ray images of the cluster.
The X-ray brightness is denoted by colors and contours.
Black dots are all sources with measured redshifts,
black filled stars are AGN classified with optical spectra, 
and open stars are AGN identified with WISE colors.
}
\label{fig_fofgrp_xray}
\end{figure*}

We use a $\delta$-test \citep{DS1988} as one of these statistical tests.
This $\delta$-test is based on the local deviations of cluster galaxies 
	from the systemic velocity ($\bar{v}$) and dispersion ($\sigma$) of the entire cluster.
For each galaxy, the deviation is defined by 
\begin{equation}
   \delta^2 = (N_\textrm{\fns nn}/\sigma^2) 
              \left[ (\bar{v}_\textrm{\fns local}- \bar{v})^2 + (\sigma_\textrm{\fns local}-\sigma)^2 \right]
\label{eqn_delta}
\end{equation}
	where $\bar{v}_\textrm{\fns local}$ and $\sigma_\textrm{\fns local}$ 
	are the local velocity mean and dispersion,
	and $N_\textrm{\fns nn}$ is the number of nearest neighbors
	that determines the range of local environment.
We use $N_\textrm{\fns nn}=11$ as suggested by \citet{DS1988}.
We calculate $\delta$ for each galaxy at 
	$|\Delta cz|/(1+z_\textrm{\fns cl})\le3000\,\textrm{km s}^{-1}$,
	$r_\textrm{\fns petro,0}\le20.5$, and $R\le30^\prime$.
In Figure \ref{fig_dsdiagram},
	we visualize $\delta$ for each galaxy
	with a circle of a radius proportional to $e^{\delta}$ (black open circles).
A larger circle means a larger deviation from the global kinematics,
	thus groups of large circles represent galaxy groups 
	that are kinematically distinguished from the host cluster.
As expected, some of the FoF groups appear as kinematically distinctive subgroups:
	some parts of Group \#1 and 2, Group \#5, 8, 13, and 14.

We also calculate the statistical significance of the presence of substructures of A2199
	using a Monte Carlo simulation of the $\Delta$ statistics of \citet{DS1988},
	where $\Delta$ is the sum of $\delta$ of all the cluster galaxies.
We first generate 500 simulated clusters
	by shuffling velocities of the cluster galaxies at their observed positions
	so that the simulated clusters are similar to the observed cluster
	except for the velocity clustering property.
We then calculate $\Delta$ for the observed and the simulated clusters,
	and compute the fraction of the simulated clusters with 
	$\Delta_\textrm{\fns sim}>\Delta_\textrm{\fns obs}$.
We obtain $f(\Delta_\textrm{\fns sim}>\Delta_\textrm{\fns obs})=11\%$,
	which suggests the existence of possible substructures in A2199.

\begin{figure*}
\centering
\includegraphics[width=0.79\textwidth]{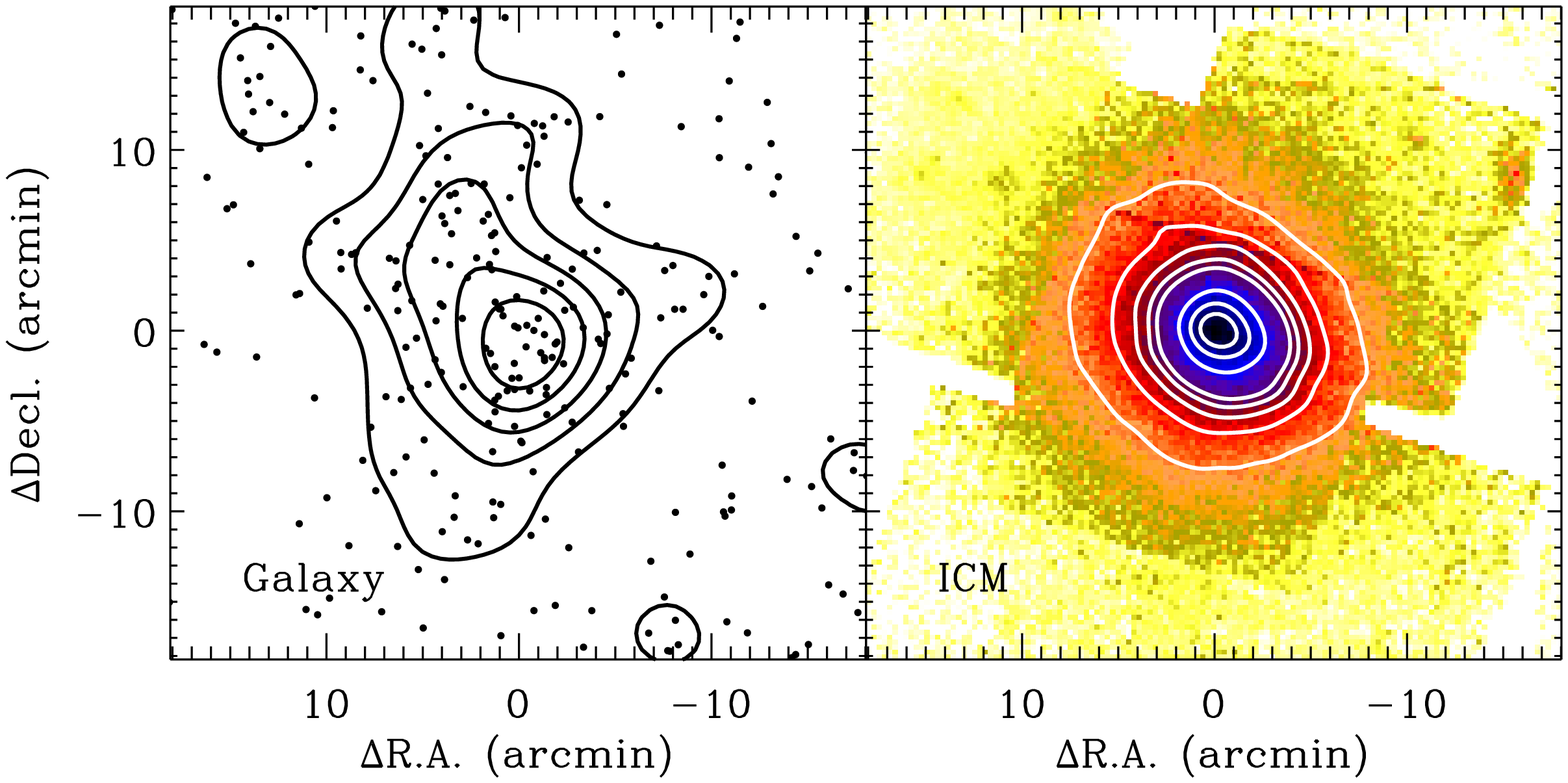}
\caption{(Left) Iso-surface number density contours of cluster galaxies (black solid lines)
and cluster galaxies (black dots). 
(Right) X-ray iso-intensity contours (white solid lines) 
with X-ray emission color-coded by its intensity.
Different color code from one in Figure \ref{fig_fofgrp_xray} is used
to better resolve the central part.
}
\label{fig_galden_xray_center}
\end{figure*}

\subsubsection{Spatial Correlation between FoF galaxy groups and X-ray emission}
We now examine the correlation of spatial distribution of the FoF galaxy groups
	with that of X-ray emission.
We use the Suzaku X-ray images of A2199 
	(T. Tamura et al. 2017, in prep.; see colors and contours in Figure \ref{fig_fofgrp_xray}).
However, the spatial coverage of the Suzaku observations is not complete
	(Figure \ref{fig_fofgrp_xray} shows several places of detector gaps),
	thus we first focus on a simple identification of X-ray counterparts of
	the optically selected galaxy groups identified in the previous section.
Figure \ref{fig_fofgrp_xray} shows the spatial distribution of the cluster galaxies on top of the X-ray intensity map.
We also overplot active galactic nuclei (AGN, star symbols).
Filled star symbols indicate the AGNs classified with optical spectra
	(i.e. $z_\textrm{\fns specclass}$ eq `QSO', $z_\textrm{\fns subclass}$ eq `AGN', 
	or $z_\textrm{\fns subclass}$ eq `BROADLINE' in the SDSS database),
	and open star symbols denote the AGNs identified with WISE colors using the criteria of \citet{Mateos_etal2012}.
Because the WISE criteria for AGNs do not require redshift information,
	we use all the WISE AGN candidates in the photometric sample regardless of redshift information.

Figure \ref{fig_fofgrp_xray} shows that 
	the strong X-ray emission in the very central region ($R<15^\prime$)
	matches well with the main body of A2199 (FoF group \#1), as expected.
To better compare the spatial distribution of the galaxies with that of the ICM in the central region,
	we show the galaxy number density contours at $R\lesssim18^\prime$ 
	in the left panel of Figure \ref{fig_galden_xray_center}
	with the X-ray intensity map and contours in the right panel.
Both contours at $R\lesssim10^\prime$ show a northeast-southwest elongation.
Figure \ref{fig_fofgrp_xray} also shows that 
	the most small X-ray clumps coincide with the AGNs,
	meaning that the AGNs are mainly responsible for the X-ray emission in those clumps.

\begin{figure*}
\centering
\includegraphics[width=\textwidth]{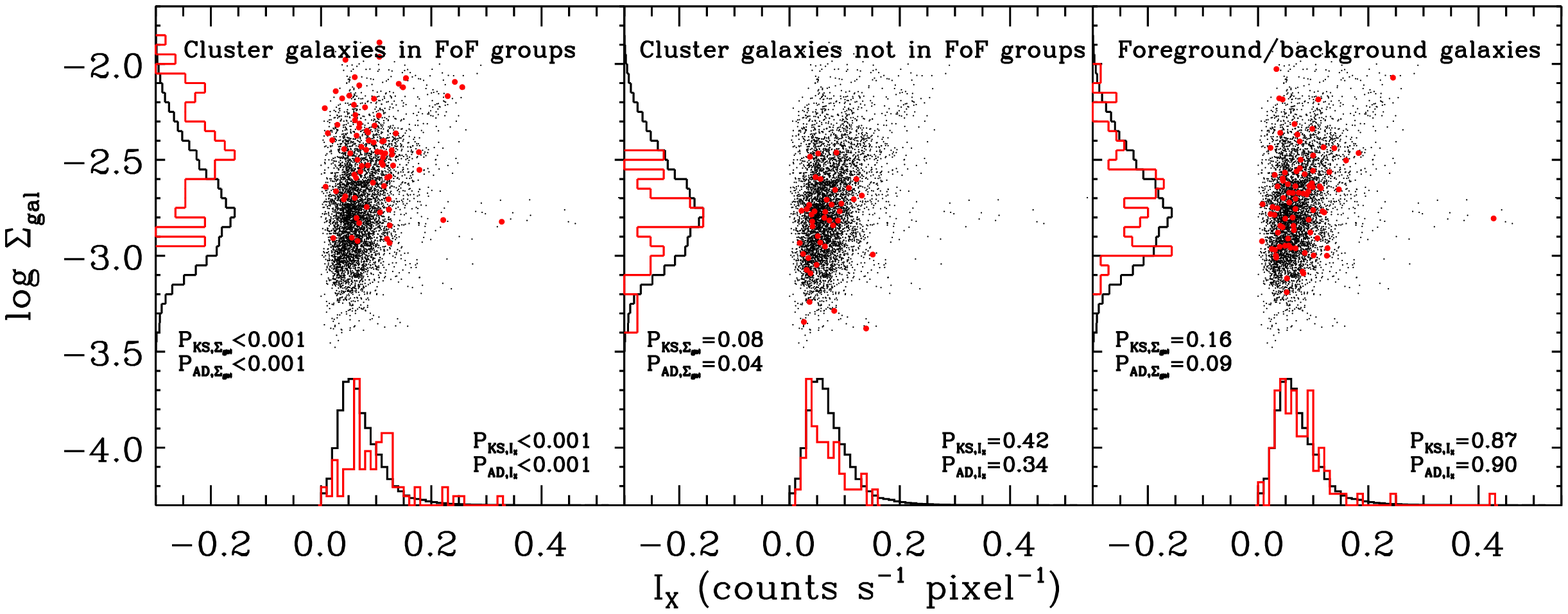}
\caption{Correlation between surface galaxy number density and X-ray intensity in A2199.
Black dots indicate the correlation between the two in the entire region where the X-ray
measurements are available (see Figure \ref{fig_fofgrp_xray}).
Red dots in each panel indicate the correlation between the two at the positions of
galaxies in three different galaxy samples: cluster galaxies in the FoF groups (left),
cluster galaxies not in the FoF groups (middle), and foreground/background galaxies (right).
Histograms along X- and Y-axes show the distributions of X-ray intensity and
the surface galaxy number density, respectively.
Black and red histograms indicate, respectively, the distributions for the entire sample
and the subsample of each panel.
Two numbers in the left and right corners indicate $p$-values from the K-S and A-D $k$-sample tests
on the distributions of the entire sample and the subsample.
}
\label{fig_galden_xray_histo}
\end{figure*}

To examine the correlation between FoF groups and X-ray emission,
	we perform the following statistical test.
We first construct the surface galaxy number density map using the cluster member galaxies at $r_\textrm{\fns petro,0}<17.77$.
We use the method in \citet{Gutermuth_etal2005} based on the distance to the fifth closest galaxy to construct the number density map,
	and the pixel scale of the map is the same as the one in the X-ray map (i.e. $16.7^{\prime\prime} \textrm{pix}^{-1}$).
The black dots in Figure \ref{fig_galden_xray_histo} show the surface number density distribution
	as a function of X-ray intensity for the entire region where the X-ray data exist (see Figure \ref{fig_fofgrp_xray}).

We then overplot the distribution of pixels at the positions of galaxies in three different galaxy samples:
	cluster galaxies in the FoF groups, cluster galaxies not in the FoF groups, and foreground/background galaxies
	(red dots in three panels).
To reduce the effect of the cluster main component, 
	we mask the central $20^\prime$ region and
	do not plot the galaxies in the main body of A2199 (i.e. FoF group \#1).
The plot shows that the surface number densities of the galaxies in the FoF groups (left panel)
	are higher than those of the other samples, as expected.
However, only the galaxies in the FoF groups show that both the galaxy number density and the X-ray intensity
	distributions are skewed to higher values compared to the distributions of the entire sample,
	indicating a correlation of the spatial distribution between FoF groups and X-ray emission.
We list the $p$-values from the K-S and the A-D $k$-sample tests for the distributions of the subsample
	(red dots) and the parent sample (black dots) in each panel, which supports our conclusion.

\begin{figure*}
\centering
\includegraphics[width=0.7\textwidth]{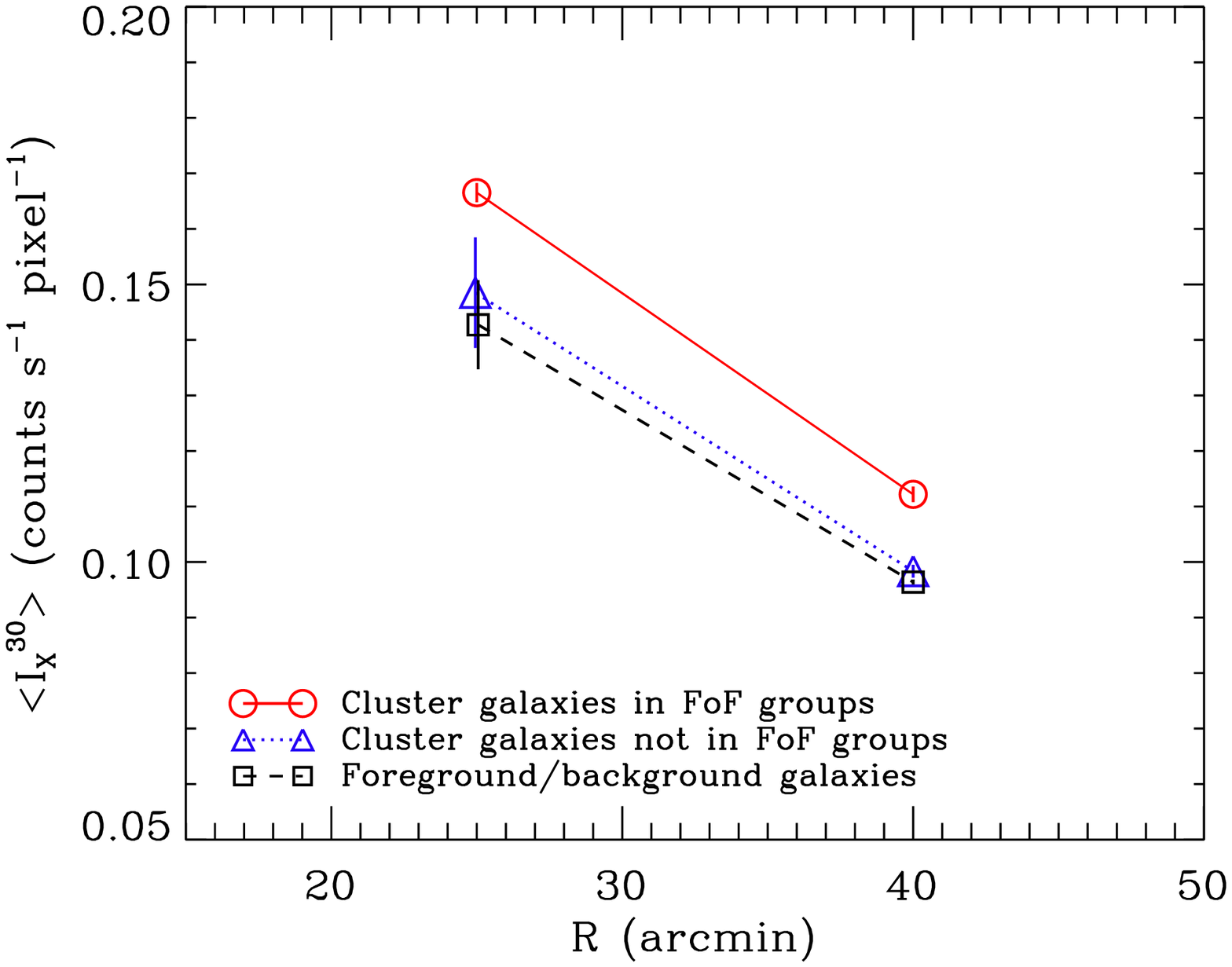}
\caption{The mean of top 30\% of the X-ray intensities measured 
at the positions of cluster galaxies in FoF groups (red circle),
at the positions of cluster galaxies not in FoF groups (blue triangle),
and at the positions of foreground/background galaxies (black square). 
The mean is measured at two different radial ranges, 
$20^\prime<R<30^\prime$ and $30^\prime<R<50^\prime$.
Error bars are calculated using the subsampling method.
}
\label{fig_xray_r}
\end{figure*}

We also perform another statistical test.
This is similar to the previous one, but we focus more on the comparison of the galaxy samples at similar radii.
Figure \ref{fig_xray_r} shows the mean of the top 30\% of the X-ray intensity distribution, $<I_X^{30}>$,
	at the positions of the galaxies in the three galaxy samples above as a function of clustercentric radius.
We plot the mean of the top 30\% of the X-ray intensity distribution
	because we are interested in the excess of X-ray intensity where galaxies are located at,
	and the conclusion does not change even though we use top 20-50\% of the X-ray intensity distribution.
To minimize the effect of the cluster main component in the X-ray map,
	we restrict our comparison to the radial ranges of $20^\prime<R<30^\prime$ and $30^\prime<R<50^\prime$.
We do not subtract the cluster main component from the X-ray map to reduce the uncertainty
	that could be introduced by the imperfect subtraction.
The plot shows that the X-ray intensities near the cluster galaxies in the FoF groups, on average,
	are higher than those of the other samples both at the two radial bins.
This again confirms the correlation of the spatial distribution between FoF groups and X-ray emission.
A detailed analysis on the comparison of galaxy distribution with the complete X-ray map
	and with the weak-lensing peaks will be a topic of future studies.

\section{CONCLUSION}
We conduct a deep, uniform redshift survey of the central region of A2199 at $R<30^\prime$.
By combining 775 new MMT/Hectospec redshifts and the data in the literature,
	we provide an extensive catalog of the A2199 galaxies 
	at $r_\textrm{\fns petro,0}<20.5$ and $R<60^\prime$,
	which is useful for future studies of this system.

We apply the caustic technique to the redshift data 
	and identify 406 member galaxies in the central region at $R<30^\prime$.
The velocity dispersion profile derived from the entire sample of member galaxies is 
	smoothly connected to the stellar velocity dispersion profile of the cD galaxy,
	and is similar to the one derived using red member galaxies.
The faint-end slope of the luminosity function at $R<30^\prime$ is 
	$\alpha=-1.26\pm0.06$, which shows no faint-end upturn feature.
The luminosity function does not change much with clustercentric radius.
These results are consistent with the previous measurement based on slightly shallower redshift surveys.

The comparison of the radial velocities between the galaxies and the ICM
	suggests that there are some regions where the velocity difference
	between the two is significant more than 2$\sigma$.
We find a hint of the rotation of the galaxies at $R<5^\prime$ around the center of A2199
	with $v_\textrm{\fns rot}=300{-}600\,\textrm{km s}^{-1}$,
	but the ICM does not show such bulk motion in the same region.
This might result from a recent minor merger event in the central region 
	suggested by \citet{Nulsen_etal2013} with the Chandra X-ray data.
To develop further understanding on merging processes undergone recently in the central region,
	more precise measurements of the ICM radial velocity and
	weak-lensing analysis for dark matter distribution are necessary.

We apply a FoF algorithm to identify galaxy subgroups,
	and identify 32 group candidates at $R<60^\prime$.
A visual comparison of the spatial distribution of the FoF groups
	with the Suzaku X-ray map suggests that
	the correspondence between the FoF groups and X-ray clumps is not obvious except the central main body.
We perform simple statistical tests on the spatial correlation between the FoF groups and X-ray emission.
The tests result in a positive correlation, indicating the physical connection between the two.
Future X-ray observations with complete spatial coverage and better spatial angular resolution
	would be useful for better comparisons between the galaxies and the ICM.
Comparisons of the galaxy distribution with the dark matter distribution from a weak-lensing analysis
	will be also valuable to have a complete picture of the formation and evolution of A2199.

\acknowledgments
We thank the anonymous referee for helpful comments.
We thank Ana Laura Serra for providing the Caustic application.
Funding for SDSS-III has been provided by the Alfred P. Sloan Foundation, the Participating Institutions, the National Science Foundation, and the U.S. Department of Energy Office of Science. The SDSS-III web site is http://www.sdss3.org/.
SDSS-III is managed by the Astrophysical Research Consortium for the Participating Institutions of the SDSS-III Collaboration including the University of Arizona, the Brazilian Participation Group, Brookhaven National Laboratory, Carnegie Mellon University, University of Florida, the French Participation Group, the German Participation Group, Harvard University, the Instituto de Astrofisica de Canarias, the Michigan State/Notre Dame/JINA Participation Group, Johns Hopkins University, Lawrence Berkeley National Laboratory, Max Planck Institute for Astrophysics, Max Planck Institute for Extraterrestrial Physics, New Mexico State University, New York University, Ohio State University, Pennsylvania State University, University of Portsmouth, Princeton University, the Spanish Participation Group, University of Tokyo, University of Utah, Vanderbilt University, University of Virginia, University of Washington, and Yale University.

{\it Facility:} \facility{MMT Hectospec}

\appendix
\section{Radial Velocity Distributions of Cluster Galaxies}
\begin{figure*}
\centering
\includegraphics[width=0.9\textwidth]{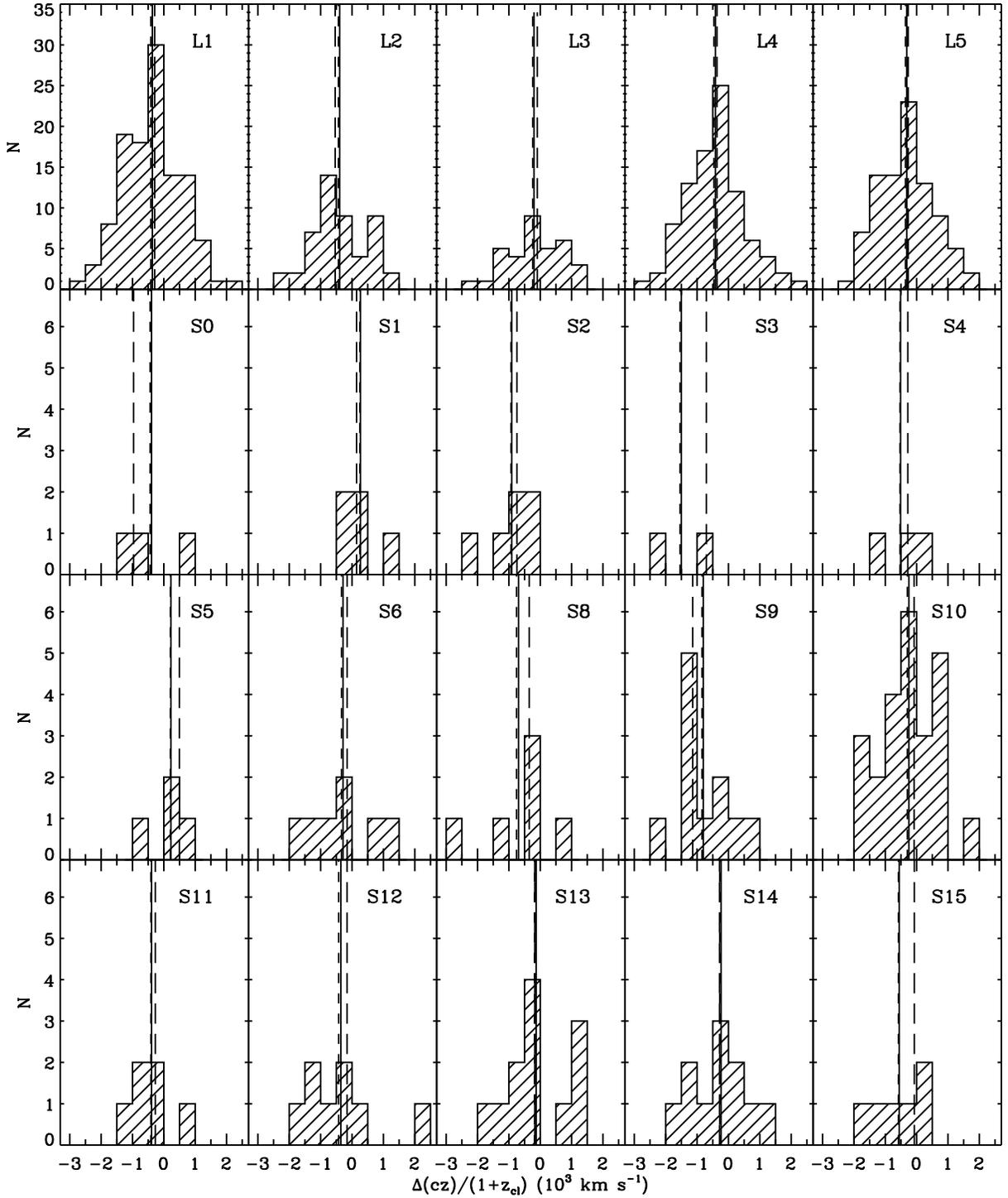}
\caption{Radial velocity distribution of cluster galaxies 
in each cell shown in Figure \ref{fig_velmap}, 
except S7 where there is only one galaxy.
The name of each cell is specified in each panel.
Mean, median, and geometric mean are denoted with
solid, long-dashed, and dashed lines, respectively.
}
\label{fig_velhist}
\end{figure*}

Here Figure \ref{fig_velhist} shows the radial velocity distribution of galaxies in each cell of Figure \ref{fig_velmap}.
We show mean, median, and geometric mean with solid, long-dashed, and dashed lines, respectively, in each panel.
The numbers of galaxies in the large and small cells are 34-91 and 1-24, respectively.
This means that we should be careful in representing the average radial velocity for the small cells.
Following the suggestion of Sauro \& Lewis (2010) that the geometric mean has less error and bias
	than the median or the mean when the sample size is less than 25,
	we could use geometric mean for the average radial velocity,
	but decide to use the mean for convenience 
	because the mean and the geometric mean agree very well in each panel.
We note that the conclusion does not change even if we use median instead of mean.

\clearpage
\bibliographystyle{apj}
\bibliography{ms}{}

\end{document}